\documentclass[10pt,journal,final]{IEEEtran}
\usepackage{mathrsfs}
\usepackage{bbding}
\usepackage{graphicx}
\usepackage{makecell}
\usepackage{float}
\usepackage{booktabs}
\usepackage[hidelinks]{hyperref}
\usepackage{cite}

\usepackage{enumitem}
\usepackage{color}
\usepackage{ulem}
\usepackage{psfrag}
\usepackage{subfigure}
\usepackage{amssymb}
\usepackage{amsthm}
\usepackage{setspace}
\usepackage{epsfig}
\usepackage{pifont}
\usepackage{amsmath}
\usepackage{array}
\newcommand{\PreserveBackslash}[1]{\let\temp=\\#1\let\\=\temp}
\newcolumntype{C}[1]{>{\PreserveBackslash\centering}p{#1}}
\newcolumntype{R}[1]{>{\PreserveBackslash\raggedleft}p{#1}}
\newcolumntype{L}[1]{>{\PreserveBackslash\raggedright}p{#1}}
\usepackage{multicol}
\usepackage{multirow}
\usepackage{diagbox}
\usepackage{pifont}
\usepackage{indentfirst}
\usepackage{amsfonts}
\usepackage{algorithm}
\usepackage{algorithmicx}
\usepackage{algpseudocode}
\floatname{algorithm}{Procedure}
\usepackage{fancyhdr}
\usepackage{amscd}
\usepackage{xcolor}
\usepackage{bm}
\usepackage{fancyhdr}
\setlist{itemsep=0pt,parsep=0pt}

\newtheorem{proposition}{Proposition}

\allowdisplaybreaks[4]

\makeatletter
\def\endthebibliography{%
	\def\@noitemerr{\@latex@warning{Empty `thebibliography' environment}}%
	\endlist
}
\makeatother
\IEEEoverridecommandlockouts

\usepackage{caption}
\begin{document}
	\title{\huge Learning Beamforming for Pinching Antenna System-Enabled ISAC in Low-Altitude Wireless Networks}
	
	% PASS, flexible antenna system, scalability, flexiblity
	
	% optimization, 
	
	\author{
		% \thanks{This work is supported by National Natural Science Foundation of China (NSFC) under Grant XXX}
		% \IEEEauthorblockN{}.
		\thanks{Jia Guo is with the School of Electronic Engineering and Computer
			Science, Queen Mary University of London, London E1 4NS, U.K. (e-mail:
			jia.guo@qmul.ac.uk).
			
			Yuanwei Liu is with the Department of Electrical and Electronic Engineering,
			The University of Hong Kong, Hong Kong (e-mail: yuanwei@hku.hk).
			
			Arumugam Nallanathan is with the School of Electronic Engineering and Computer
			Science, Queen Mary University of London, London E1 4NS, U.K. (e-mail:
			a.nallanathan@qmul.ac.uk).
		}
		\IEEEauthorblockN{Jia Guo, Yuanwei Liu, \emph{Fellow, IEEE}, and Arumugam Nallanathan, \emph{Fellow, IEEE}\vspace{-5mm}}
		
		%\IEEEauthorblockA{Queen Mary University of London\\ \{jia.guo, yuanwei.liu, a.nallanathan\}@qmul.ac.uk} 
	}
	\maketitle
	\setcounter{page}{1}
	\thispagestyle{empty}
	
	\begin{abstract}
		This work investigates the joint learning of pinching antenna (PA) positions and transmit beamforming for PA-aided integrated sensing and communication (ISAC) in the low-altitude wireless networks. By freely deploying antenna positions along waveguides, the pinching antenna system effectively mitigates the impact of path loss and thus enhances the capacities of sensing and communicating unmanned aerial vehicles (UAVs) that fly over a large range.
		We first model the problem of maximizing the sensing performance of multiple targets while satisfying the communication performance requirements of multiple users, where both the targets and users are UAVs. For mitigating in-waveguide attenuation and improving sensing performance, the segmented waveguide-enabled pinching antenna (SWAN) system is adopted. 
		Furthermore, an alternative optimization (AO) algorithm for SWAN-based ISAC (SWISAC-AO) is developed, where the optimal structure of the transmit beamforming solution is derived. A graph neural network (GNN), termed SWISAC-GNN, is then proposed to jointly learn PA positions and transmit beamforming, with its alternative update procedure inspired by the SWISAC-AO algorithm. Numerical results show that the GNN achieves sensing performance comparable to or better than the AO algorithm while better satisfying communication requirements. Moreover, the SWISAC-GNN is with much lower implementation complexity, enabling real-time deployment.

		\begin{IEEEkeywords}
			Beamforming, deep learning, integrated sensing and communications, low-altitude wireless networks, pinching antenna systems
		\end{IEEEkeywords}
	\end{abstract}

	\section{Introduction}\label{sec:intro}
%     To accommodate the rapidly increasing demand for wireless connectivity, unmanned aerial vehicles (UAVs) have been incorporated into wireless communication networks. UAVs are expected to play a pivotal role in extending human activities from terrestrial to low-altitude domains, thereby facilitating the development of the low-altitude economy (LAE).

% Integrated sensing and communication (ISAC) has been identified as one of the six key use cases for sixth-generation (6G) mobile networks, aiming to enhance network functionality and enable situational awareness of the physical environment. The integration of ISAC and UAV technologies has emerged as an efficient paradigm to support LAE \cite{isac-lae}. 

To accommodate the rapidly increasing demand for wireless connectivity, unmanned aerial vehicles (UAVs) have been incorporated into wireless communication networks and are envisioned to support emerging low-altitude applications. UAVs are expected to play a pivotal role in extending human activities from terrestrial to low-altitude domains, thereby facilitating the development of the low-altitude wireless networks (LAWNs). Integrated sensing and communication (ISAC), identified as one of the six key use cases for sixth-generation (6G) mobile networks \cite{UAV-1}, offers a unified framework that enhances network functionality and enables situational awareness of the physical environment. The integration of ISAC and UAV technologies has thus emerged as an effective paradigm to support LAWN \cite{isac-lae}.

In ISAC-enabled networks, UAVs can serve in two complementary roles: as aerial anchors providing sensing capabilities or as sensing targets to be detected and tracked \cite{UAV-1}.
In the former case, the UAV trajectory can be jointly optimized with communication and sensing strategies such as waveform design, beamforming, and power allocation to improve overall ISAC performance \cite{mu2023uav}.
In the latter case, base stations (BSs) must communicate with or sense UAVs moving over extensive areas. However, line-of-sight (LoS) blockages and severe path loss in long-distance propagation lead to notable degradation in both communication and sensing performance in conventional multi-input multi-output (MIMO) and massive MIMO systems.

To mitigate LoS blockage and free-space attenuation, various flexible antenna technologies have been proposed, including reconfigurable intelligent surfaces (RISs), fluid antennas, and movable antennas \cite{FAS-survey}. These approaches are capable of dynamically reconfiguring the propagation environment or adjusting antenna positions and shapes to enhance link quality. Nevertheless, their practical deployment is often constrained by high fabrication cost, implementation complexity, and limited spatial flexibility, which restrict their ability to effectively compensate for large-scale path loss \cite{ding2024flexible}.

Recently, the pinching antenna system (PASS) has been introduced as a low-cost and easily implementable alternative to address the aforementioned challenges. Originally proposed and prototyped by NTT DOCOMO in 2022 \cite{docomo}, PASS employs a dielectric waveguide equipped with multiple pinching antennas (PAs), which are small dielectric elements that extract signals from the waveguide in a near-wired manner and radiate them into free space. The positions of PAs can be flexibly reconfigured over a wide range, enabling a significant reduction of free-space path loss and the establishment of reliable LoS links. Theoretical analysis and experimental demonstrations have confirmed that PASS can substantially improve spectral efficiency (SE) \cite{mimo-pass} and reduce power consumption \cite{9-PA-BFOpt}.

Encouraged by its success in communication systems, PASS has been recently investigated for ISAC applications \cite{mao2025multi, pass-isac-2, 7-PA-PowerAllo-ISAC}. In such systems, beamforming plays a fundamental role in enhancing both the communication throughput and the sensing signal-to-noise ratio (SNR) \cite{mao2025multi}. The joint optimization of beamforming vectors and PA positions involves heterogeneous variables that usually require alternative optimization (AO), resulting in high computational complexity. This complexity poses a significant challenge to real-time ISAC operation in UAV-assisted scenarios. Moreover, in uplink reception, to prevent the inter-antenna radiation (IAR) effect in which signals received by one PA re-radiate through others, only one PA should be placed or activated on each waveguide \cite{ouyang2025uplink}. This restriction further limits the sensing capability of PASS, particularly when UAVs move across wide spatial regions.

To enable real-time and low-complexity decision-making, deep learning (DL) has recently emerged as a promising framework that learns wireless policies offline. A policy can be viewed as a mapping from environmental features, such as channel matrices, to control decisions, such as beamforming weights. DL-based approaches have demonstrated significant advantages in enabling real-time inference, joint optimization with channel acquisition, and robustness against various impairments including channel estimation errors. Several studies have already introduced DL for beamforming design in PASS \cite{guo2025gpass, meta-pass}. Nevertheless, the design of deep neural networks (DNNs) that can efficiently learn beamforming strategies for ISAC in PASS-aided UAV systems with adaptability to varying environments (say number of users) remains an open and challenging research problem.
    
    \subsection{Related Works}
    \subsubsection{Optimizing Beamforming for PASS}
    In PASS, the positions of PAs (also called pinching beamforming) have been jointly optimized with transmit beamforming to minimize power consumption \cite{pass-joint,pass-pow-radi-xxx,pass-noma-gdq, 9-PA-BFOpt} or maximize SE \cite{pass-in-wg,pass-element-wise-opt,pass-noma-se,mimo-pass}. 
    In \cite{pass-joint}, the power consumption of moving PAs was considered in the system model and optimization problem. A power radiation model was proposed in \cite{pass-pow-radi-xxx} such that the power leaked from every PA can be controlled and optimized.

    To jointly optimize the PA positions and transmit beamforming matrices that are with different modalities, AO is usually adopted. The PA positions can be optimized via one-dimensional search \cite{mimo-pass, pass-element-wise-opt,9-PA-BFOpt}, intelligent optimization algorithms such as particle swarm optimization \cite{pass-in-wg} or branch-and-bound algorithm (for discrete activation of PAs) \cite{pass-pow-radi-xxx}. The transmit beamforming matrix can be optimized via block coordinate descent \cite{pass-joint, pass-in-wg}, fractional programming \cite{mimo-pass} or finding the optimal solution structure \cite{pass-pow-radi-xxx}. Suboptimal beamforming with closed-form expressions such as zero-forcing (ZF) and minimum-mean-square-error (MMSE) beamformers can also be adopted for lower computational complexity \cite{pass-element-wise-opt,9-PA-BFOpt}.
    
    These existing works focused on maximizing the communication performance. When it comes to ISAC systems, the positions of PAs and the beamformers need to be optimized to guarantee both the communication and sensing performance. How to optimize them jointly with affordable computational complexity is still under-investigated.

    \subsubsection{PASS-aided ISAC}
    PASS has been recently introduced to support ISAC. The rate region of the PASS-aided ISAC system in a single-user single-target scenario was derived in \cite{ouyang2025rate}. 
    The closed-form Cramer-Rao bound (CRB) was derived in \cite{pa-isac-6} for establishing the fundamental sensing limits of PASS.
    In \cite{6-PA-downlinkBF}, it was derived that the CRB achieved by PASS is much lower than the one achieved by conventional antennas. 
    For sensing in PASS, a metric called Bayesian Cramer-Rao bound was proposed in \cite{pass-bcrb}, which is independent of the exact values of sensing parameters (say target positions).
    
    The PA activation was optimized in \cite{pass-isac-2} via successive convex optimization (SCA) to minimize the outage probability of sensing while satisfying a transmission data rate constraint. In \cite{mao2025multi}, the PA positions and beamforming was jointly optimized in a single-user single-target scenario to maximize communication rate under a radar SNR constraint. 

    Machine learning was also introduced into PASS-aided ISAC for optimization. In \cite{pass-merl}, a maximum entropy-based reinforcement learning algorithm was designed to maximize SE while satisfying a sensing SNR constraint.

    When there are multiple PAs on a waveguide for receiving signals, the IAR effect may occur, i.e., signals captured by one PA re-radiate through other PAs. To avoid this, existing works either adopted the setting that there is only one PA on every waveguide \cite{mao2025multi}, or assumed that there is only one PA activated in every time slot \cite{pass-isac-2}. This can somewhat restrict the sensing performance of PASS. Besides, most existing works considered rather simple settings such as single target \cite{mao2025multi} or single waveguide \cite{pass-merl}. When it comes to more practical systems such as multiple waveguides, PAs, targets and users, the beamforming optimization may incur high computational complexity.
    
    \subsubsection{Graph Neural Networks for Wireless Communications}
    DNNs have been widely applied to learning wireless resource allocation policies. Among the DNN architectures, graph neural networks (GNNs) are more efficient than fully-connected neural networks (FNNs) and convolutional neural networks in better size-generalizability \cite{DYX,Eisen2020,shen2019graph} and scalability \cite{DYX,lee2020wireless,GJ_TWC_GNN,GNN-PC-CellFree-TWC2024}, which indicates that the GNNs still perform well under unseen problem scales (say the number of users) without re-training, and the complexity for training GNNs in large-scale problems is low. The advantages of GNNs stem from the ability to satisfy permutation properties widely exist in wireless communications. 

    The permutation property comes from sets in a problem. For example, in a beamforming optimization problem in a multi-user multi-input single output system, the antennas and the users constitute two sets. The beamforming policy satisfies the two-dimensional permutation equivariant property, i.e., the beamforming matrix is not affected by changing the orders of antennas and users \cite{ZBC_WCNC}. For learning a policy with GNNs, it is important that the permutation property satisfied by the GNN matches that of the policy. Otherwise, the size-generalizability or scalability would be deteriorated \cite{GJ_TWC_GNN}. To exhibit matched permutation property to a policy, a framework of identifying sets from a problem, modeling graphs and designing parameter sharing of GNNs was proposed in \cite{LSJ}.

    It has been widely believed that the GNNs are naturally size-generalizable and scalable, because parameter sharing is introduced into GNNs to satisfy permutation properties \cite{LSJ}. Nonetheless, it has recently been found in \cite{GJ-RGNN} that the size generalizability and scalability of a GNN depend on its processor, which is a parameterized function for extracting information from neighboring vertices and edges. The analysis results in \cite{guo2025attention} further show that for learning policies whose environmental parameters do not reflect interference (say beamforming), the interference should be modeled in the processor with an attention mechanism.
    
    \subsection{Motivations and Contributions}
    Motivated by the success of PASS for serving users and sensing targets with the mitigated impact of path loss, we strive to investigate PASS-aided ISAC in a UAV system for supporting LAWN, where the users and targets are UAVs. To serve the UAVs that fly over a large range, long waveguides need to be deployed, such that in-waveguide attenuation is not negligible \cite{pass-in-wg}. To mitigate this impact, the segmented waveguide-enabled PASS (SWAN) proposed in \cite{ouyang2025uplink} is adopted, where each waveguide is divided into multiple segments, each with an individual feed point. This structure also enables multiple receiving PAs on each waveguide without incurring the IAR effect, thereby improving sensing performance. 
    We first propose a numerical algorithm for optimizing beamforming in this PASS-aided UAV system for ISAC. 
    Motivated by the ability of real-time implementation of DNNs, the size-generalizability and scalability of GNNs, we then propose a GNN to learn the beamforming policy.
    
    The main contributions of this paper are summarized as follows.
    \begin{itemize}
        \item We formulate the beamforming optimization problem for multi-target multi-user ISAC in a PASS-aided UAV system. The problem aims to maximize the sensing rate (SR) of targets while satisfying a quality-of-service (QoS) constraint for users. The SWAN structure is considered, which can mitigate the impact of in-waveguide loss and improve the performance of sensing. 
        \item We transform the highly non-convex problem to one that is convex with respect to (w.r.t.) some of the variables when fixing other variables. Then,
        we propose an \textbf{AO} algorithm for \textbf{SW}AN-based \textbf{ISAC} (SWISAC-AO) to solve the optimization problem. We derive the optimal solution structure of the transmit beamforming, with which the high-dimensional beamforming matrix only depends on some low-dimensional variables. 
        \item We propose a GNN called SWISAC-GNN for learning the beamforming policy. The update equation of the GNN is inspired by the SWISAC-AO algorithm, particularly in (i) the alternating update procedure and (ii) the mechanism of information aggregation from neighboring vertices and edges. The proposed GNN effectively captures the multi-user interference (MUI), which is crucial for learning beamforming. The derived optimal beamforming structure is also leveraged to simplify the mappings to be learned. 
        \item The performance of the proposed SWISAC-AO algorithm and the SWISAC-GNN is validated via simulations. It can be seen that the SWISAC-GNN can achieve SR close to or better than the SWISAC-AO algorithm, but the ability to satisfy QoS constraints is better than that of the algorithm. Moreover, the SWISAC-GNN is computationally efficient for implementation in real-time, and can be fast-adapted to different problem scales.
    \end{itemize}

    While a part of this work has been published in \cite{GJ-GC25}, the contents in this journal version are substantially extended, including proposing the SWISAC-AO algorithm for solving the beamforming optimization problem, deriving the optimal solution structure of transmit beamforming, and adopting the SWAN structure. 

    % $(\mathbf{A})_{kj}$ denotes the element on the $k$-th row and the $j$-th column of the matrix $\mathbf{A}$.

	\section{System Model and Problem Formulation} \label{sec:problem-policy}
	
\begin{figure}[!htb]
		\centering
		\includegraphics[width=\linewidth]{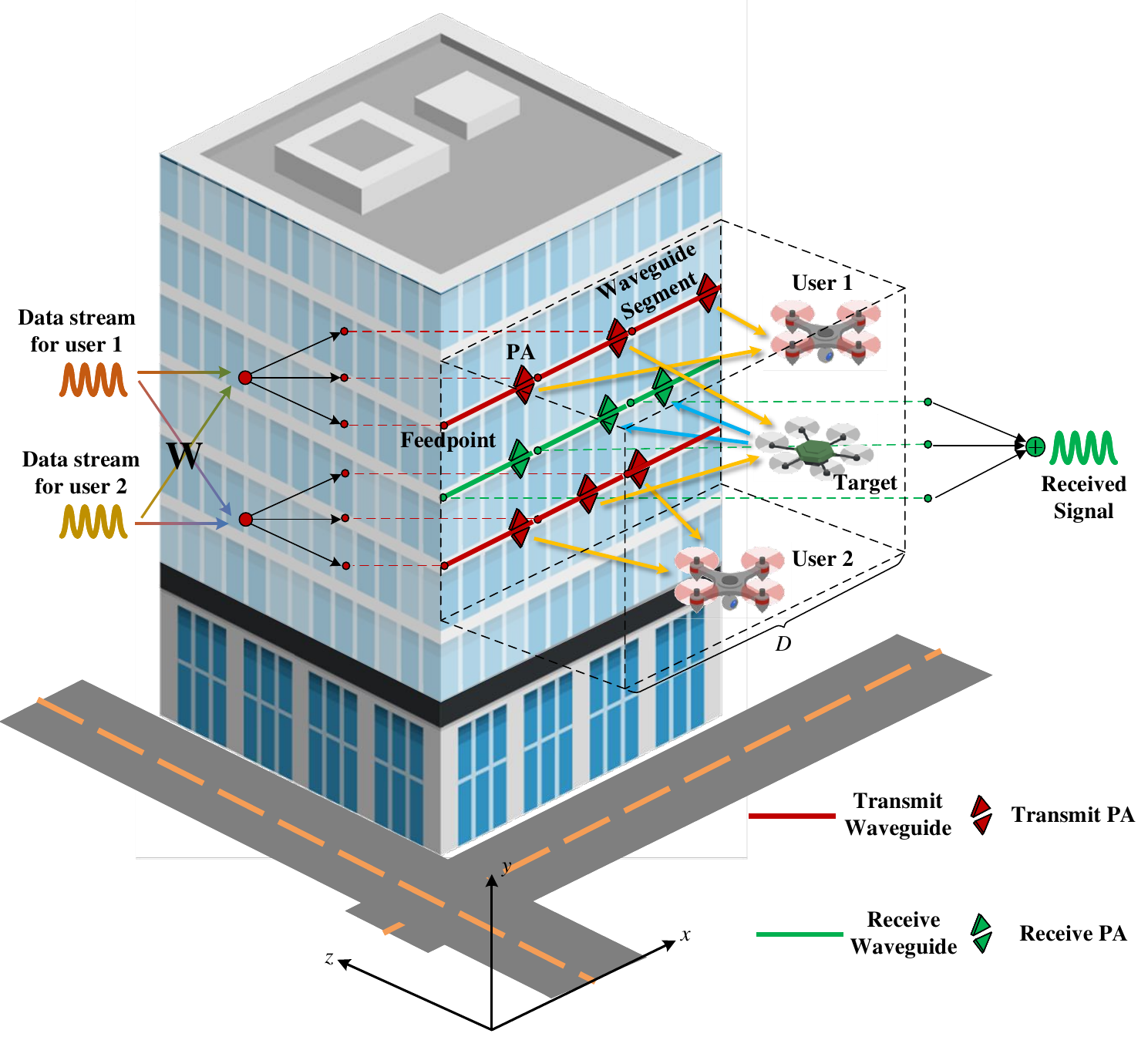}
		\caption{System model, $N_{\sf t}=2,N_{\sf r}=1, S=3, K_{\sf c}=2, K_{\sf s}=1.$}
		\label{fig:sysmodel}
	\end{figure}

	Consider an UAV-aided ISAC shown in Fig. \ref{fig:sysmodel}, where a BS deployed on a building transmits to $K_{\sf c}$ single-antenna users and senses $K_{\sf s}$ targets. Both the users and the targets are UAVs. They are deployed in a cube with the $x$-axis side length being $D$.
	There are $N_{\sf t}$ dielectric waveguides deployed on the building for transmission and $N_{\sf r}$ waveguides for reception.
	Without the loss of generality, it is assumed that the waveguides are parallel to the $x$-axis.  The length of each waveguide is the same as the side of the square area.
	
	To mitigate the impact of in-waveguide attenuation and the IAR effect, we consider the SWAN structure proposed in \cite{ouyang2025uplink}. Specifically, each waveguide is composed of $S$ short segments arranged end-to-end. The segments are not physically connected, and each segment has its own feed point for injecting signals into the segment. 
	
	We adopt the segment aggregation protocol in the uplink and the segment selection protocol in the downlink. To be more specific, in the uplink, signals extracted from all segments are aggregated and forwarded to the RF chain for baseband processing. In the downlink, only a single selected segment is connected to the RF chain.
	
	There are $M$ PAs deployed every transmission waveguide segment and one PA deployed on each receive waveguide segment. %The segmented structure also enables the receiving waveguide to receive signals by multiple PAs without incurring the inter-antenna radiation, i.e., the signal captured by a PA is re-radiated from other PAs.
	Denote the coodinate of the $m$-th PA on the $s$-th segment of the $n$-th transmission waveguide as $\bm{\phi}_{{\sf t}, msn}=[x_{{\sf t},msn}, y_{{\sf t},n}, z_{{\sf t},n}]^\mathsf{T}$. 
	Denote the coordinate of the PA on the $s$-th segment of the $n$-th receiving waveguide as $\bm{\phi}_{{\sf r}, sn}=[x_{{\sf r},sn}, y_{{\sf r},n}, z_{{\sf r},n}]^\mathsf{T}$.
	
	Denote $\mathbf{s}_k=[s_{k1},\cdots,s_{kL}]^T \in {\mathbb C}^{L}$ as the signal transmitted to the $k$-th user, where $L$ is the length of communication frame or the number of sensing pulses, $\|\mathbf{s}_k\|=L$. As most applications of ISAC for PASS are for scenarios with high-frequency bands where LoS propagation dominates, we consider a free-space LoS channel model in the sequel.
	
	\subsection{Communication Model}
	The received signal at the $k$-th user can be expressed as,
	\begin{align}\label{eq:receive-sig}
		\mathbf{y}_k^\mathsf{T} = &\mathbf{h}^\mathsf{H}(\mathbf{u}_k, \mathbf{\Phi}_{\sf t})\mathbf{G}(\mathbf{\Phi}_{\sf t})\mathbf{w}_k\mathbf{s}_k^{\mathsf{T}} + \notag\\
		&\textstyle\sum_{j=1,j\neq k}^{K_{\sf c}} \mathbf{h}^\mathsf{H}(\mathbf{u}_k, \mathbf{\Phi}_{\sf t})\mathbf{G}(\mathbf{\Phi}_{\sf t})\mathbf{w}_j\mathbf{s}_j^{\mathsf{T}} + \mathbf{n}_k^{\mathsf{T}},
	\end{align}
	where $\mathbf{u}_k$ denotes the position of the $k$-th user, $\bm{\Phi}_{\sf t}=[\bm{\phi}_{{\sf t}, 111},\cdots,\bm{\phi}_{{\sf t}, MSN_{\sf t}}]$ contains the positions of all the transmitting PAs, $\mathbf{h}(\mathbf{u}_k, \mathbf{\Phi}_{\sf t})=[h(\mathbf{u}_k, \bm{\phi}_{{\sf t}, 111}),\cdots,h(\mathbf{u}_k, \bm{\phi}_{{\sf t},MSN_{\sf t}})]^\mathsf{T}$ is the free-space channel vector from the transmitting PAs to the $k$-th user, and
	\begin{align}\label{eq:chl}
		h(\mathbf{u}_k, \bm{\phi}_{{\sf t},msn}) = \frac{\eta^{\frac{1}{2}}e^{-jk_0 \|\mathbf{u}_k-\bm{\phi}_{{\sf t},msn}\|}}{\|\mathbf{u}_k-\bm{\phi}_{{\sf t},msn}\|},
	\end{align}
	$\eta\triangleq\frac{c^2}{16\pi^2 f_c^2}$, $c$ and $f_c$ respectively denote the speed of light and the carrier frequency, $k_0=\frac{2\pi}{\lambda}$ is the wavenumber, and $\lambda$ is the free-space wavelength, $\mathbf{n}_k$ is the noise vector with $\mathbf{n}_k\sim\mathcal{CN}(\mathbf{0}, \sigma_{\sf c}^2\mathbf{I}_L)$, $\mathbf{w}_k$ is the transmit beamforming vector for the $k$-th user. Moreover, $\mathbf{G}(\mathbf{\Phi}_{\sf t})={\rm diag}(\mathbf{g}(\mathbf{\Phi}_{{\sf t},1}),\cdots,\mathbf{g}(\mathbf{\Phi}_{{\sf t},N_{\sf t}}))$ is the in-waveguide channel matrix,
	%%	 that can be expressed as,
	%%	\begin{align}
	%%		\mathbf{G}(\mathbf{\Phi}_{\sf t})=\begin{bmatrix}
	%%			\mathbf{g}_1(\mathbf{\Phi}_{t1}) & \cdots & \mathbf{0} \\
	%%			\vdots & \ddots & \vdots \\
	%%			\mathbf{0} & \cdots & \mathbf{g}_N(\mathbf{\Phi}_{tN}),
	%%		\end{bmatrix}
	%%	\end{align}
	where $\mathbf{g}(\mathbf{\Phi}_{{\sf t},n})=[\mathbf{g}(\mathbf{\Phi}_{{\sf t},1n})^{\sf T},\cdots,\mathbf{g}(\mathbf{\Phi}_{{\sf t},Sn})^{\sf T}]^{\sf T}$, $\bm\Phi_{{\sf t},sn}=[\bm\phi_{{\sf t},1sn},\cdots,\bm\phi_{{\sf t},Msn}]$,
	\begin{align}\label{eq:in-wg-chl}
		\mathbf{g}(\mathbf{\Phi}_{{\sf t},sn}) &\!=\! \frac{1}{\sqrt{M}}\!\Big[
		e^{-j\frac{2\pi}{\lambda_g}\|\bm{\phi}_{{\sf t},0sn}\!-\!\bm{\phi}_{{\sf t},1sn}\|}\!\cdot\! \underbrace{10^{-\frac{\kappa}{20}\|\bm{\phi}_{{\sf t},0sn}\!-\!\bm{\phi}_{{\sf t},1sn}\|}}_{\text{In-waveguide loss}},\notag\\
		&\hspace{-5mm}\!\cdots,\!
		e^{-j\frac{2\pi}{\lambda_g}\|\bm{\phi}_{{\sf t},0sn}-\bm{\phi}_{{\sf t},Msn}\|}\cdot \underbrace{10^{-\frac{\kappa}{20}\|\bm{\phi}_{{\sf t},0sn}-\bm{\phi}_{{\sf t},Msn}\|}}_{\text{In-waveguide loss}}
		\Big]^{\sf T}\!\!
	\end{align}
	is the in-waveguide channel vector on the $s$-th segment of the $n$-th waveguide if the segment is selected, and $\mathbf{g}(\mathbf{\Phi}_{{\sf t},sn})=\mathbf{0}$ otherwise, $\bm{\phi}_{{\sf t},0sn}$ is the position of the feed-point of the $s$-th segment of the $n$-th transmitting waveguide, $\lambda_g=\lambda/n_{\rm eff}$ is the guided wavelength with $n_{\rm eff}$ being the effective refractive index of the dielectric waveguide. $\kappa$ denotes the average attenuation factor along the dielectric waveguide in dB/m. 
	The in-waveguide loss grows with the distance from the feeding point to the PA.
	By using the segmented structure, each segment is short such that the in-waveguide loss is small. Hence, we omit this loss in the following derivations.
	
	From \eqref{eq:receive-sig}, the communication rate (CR) of the $k$-th user is given by,
	\begin{align}
		&R_k(\mathbf{U}, \bm{\Phi}_t) =\notag\\
		&\log_2\Bigg(1+\frac{|\mathbf{h}^{\sf H}(\mathbf{u}_k, \bm{\Phi}_{\sf t})\mathbf{G}(\bm{\Phi}_{\sf t})\mathbf{w}_k|^2}{\sum_{j=1,j\neq k}^{K_{\sf c}} |\mathbf{h}^{\sf H}(\mathbf{u}_k, \bm{\Phi}_{\sf t})\mathbf{G}(\bm{\Phi}_{\sf t})\mathbf{w}_j|^2 + \sigma_{\sf c}^2}\Bigg),
	\end{align}
	where $\mathbf{U}=[\mathbf{u}_1,\cdots,\mathbf{u}_{K_{\sf c}}]$.
	
	\subsection{Sensing Model}
	The signals transmitted from the transmit PAs are reflected by the targets and then received by the receiving PAs. The received signal can be expressed as,
	\begin{align}
		\mathbf{Y}_{\mathsf{r}} = \sum_{k=1}^{K_{\sf s}}\beta \mathbf{G}(\bm{\Phi}_{\sf r}) \mathbf{h}(\mathbf{u}_{{\sf tar},k}, \bm{\Phi}_{\sf r})\mathbf{h}^{\sf H}(\mathbf{u}_{{\sf tar}, k}, \mathbf{\Phi}_{\sf t} )\mathbf{G}(\mathbf{\Phi}_{\sf t} )\mathbf{W}\mathbf{S} \!+\! \mathbf{N}_{\sf s}, \notag
	\end{align}
	where $\beta\sim\mathcal{CN}(0, \alpha_s)$ is the radar cross-section of the $k$-th target, $\bm{\Phi}_{\sf r}=[\bm{\phi}_{{\sf r},11},\cdots,\bm{\phi}_{{\sf r},SN_{\sf r}}]$ contains the positions of all the receiving PAs, $\mathbf{G}(\bm{\Phi}_{\sf r})={\rm diag}(\mathbf{g}(\bm{\Phi}_{{\sf r},1}),\cdots,\mathbf{g}(\bm{\Phi}_{{\sf r},N_{\sf r}}))$, $\mathbf{g}(\bm{\Phi}_{{\sf r},n})=[g(\bm{\Phi}_{{\sf r},1n}),\cdots,g(\bm{\Phi}_{{\sf r},Sn})]$, $g(\bm{\Phi}_{{\sf r},sn})=e^{-j\frac{2\pi}{\lambda_g}\|\bm{\phi}_{{\sf r},0sn}-\bm{\phi}_{{\sf r},sn}\|}$, $\bm{\phi}_{{\sf r},0Sn}$ is the feeding point of the $n$-th receiving waveguide, $\mathbf{u}_{{\sf tar},k}$ is the position of the $k$-th target,
	$\mathbf{h}(\mathbf{u}_{{\sf tar},k}, \bm{\Phi}_{\sf r})=[h({\mathbf{u}_{{\sf tar},k}}, \bm{\phi}_{{\sf r},11}),\cdots,h({\mathbf{u}_{{\sf tar},k}}, \bm{\phi}_{{\sf r},SN_{\sf r}})]$,
	$\mathbf{S}=[\mathbf{s}_1,\cdots,\mathbf{s}_K]^{\sf T}$, ${\rm vec}(\mathbf{N}_{\sf s})\sim\mathcal{CN}(\mathbf{0}, \sigma_{\sf s}^2\mathbf{I}_{N_{\sf r} L})$ is the noise vector, ${\rm vec}(\cdot)$ denotes the operation of concatenating each column of a matrix as a vector.
	
	The information-theoretic limit on this sensing task is characterized by the sensing mutual information (MI), while SR characterizes the sensing MI per unit time. 
	
	According to \cite{oy-isac}, the SR can be expressed as,
	\begin{align}
		&R_{\sf s}(\bm{\Phi}_{\sf t},\mathbf{U}_{\sf tar}, \bm{\Phi}_{\sf r}) = \notag\\
		&\sum_{n=1}^{N_{\sf r}} \log_2\Bigg(1+L\frac{\alpha\|\tilde{\mathbf{h}}^{\sf H}(\bm{\Phi}_{\sf t},\mathbf{U}_{\sf tar}, \bm{\phi}_{{\sf r},n})\mathbf{W}\|^2}{\sigma_{\sf s}^2}\Bigg),
	\end{align}
	where $\tilde{\mathbf{h}}^{\sf H}(\bm{\Phi}_{\sf t},\mathbf{U}_{\sf tar}, \bm{\phi}_{{\sf r},n})\triangleq \sum_{k=1}^{K_{\sf s}}\sum_{s=1}^S g(\bm{\phi}_{{\sf r},sn})h({\mathbf{u}_{{\sf tar},k}},$ $ \bm{\phi}_{{\sf r},sn})\mathbf{h}^{\sf H}({\mathbf{u}_{{\sf tar},k}}, \bm{\Phi}_{{\sf t}})\mathbf{G}({\bm{\Phi}_{\sf t}})$, $\mathbf{U}_{\sf tar} = [\mathbf{u}_{{\sf tar},1},\cdots,\mathbf{u}_{{\sf tar},K_{\sf s}}]$, $\mathbf{W}=[\mathbf{w}_1,\cdots,\mathbf{w}_{K_{\sf c}}]$.
 	
 	\subsection{Beamforming Optimization Problem Formulation}
 	We consider jointly optimizing the transmit beamforming and the PA positions (called pinching beamforming in the sequel) to maximize the SR while satisfying the minimum CR constraints for guaranteeing the QoS of the users. 
 	
 	The optimization problem can be formulated as,
 	\begin{subequations}\label{eq:prob}
 		\begin{align}
 			\max_{\bm{\Phi}_{\sf t}, \bm\Phi_{\sf r}, \mathbf{W}} & R_{\sf s}(\bm{\Phi}_{\sf t},\mathbf{U}_{\sf  tar}, \bm{\Phi}_{\sf r}) \label{eq:obj0} \\
 			{\rm s.t.} & R_k(\mathbf{U}, \bm{\Phi}_t) \geq R_{\min}, \forall k = 1,\cdots,K_{\sf c}, \label{eq:const-r}\\
 			& 0 \leq x_{{\sf t}, msn}- x_{{\sf t}, 0sn} \leq \frac{D}{S}, \forall m,n, \label{eq:const-tx}\\
 			& 0 \leq x_{{\sf r}, sn}- x_{{\sf r}, 0sn} \leq \frac{D}{S}, \forall n,\label{eq:const-rx}\\
 			& x_{{\sf t}, msn} - x_{{\sf t}, (m-1)sn} \geq \Delta_{\min}, \forall m,n \label{eq:p-cons-min-d}\\
 			& \|\mathbf{W}\|_F^2 \leq P_{\max}, \label{eq:p-cons-power}
 		\end{align}
 	\end{subequations}
 	where $R_{\min}$ is the minimum CR requirement, \eqref{eq:p-cons-min-d} restricts the minimum distance among PAs on every transmitting waveguide to avoid mutual coupling, and \eqref{eq:p-cons-power} is the constraint that the transmit power cannot exceed a power budget $P_{\max}$.
 	
 	The joint transmit and pinching beamforming policy is the mapping from the known parameters to the optimized variables of problem \eqref{eq:prob}, which can be expressed as $\{\bm{\Phi}_{\sf t}, \bm\Phi_{\sf r}, \mathbf{W}\}=F_{\sf BF}(\mathbf{U},\mathbf{U}_{\sf tar})$.
 	
 	By finding permutations that do not affect the objective function and constraints of problem \eqref{eq:prob}, it is not hard to prove that the beamforming policy satisfies the following permutation property,
 	\begin{align}\label{eq:pe}
 		\{\bm{\Phi}_{\sf t}\bm{\Omega}_{\sf t}, \bm\Phi_{\sf r}\bm{\Omega}_{\sf r}, \bm{\Pi}_{\sf t}^{\sf T}\mathbf{W}\bm{\Pi}_{\sf ue}\}=F_{\sf BF}(\mathbf{U}\bm{\Pi}_{\sf ue},\mathbf{U}_{\sf tar}\bm{\Pi}_{\sf tar}),
 	\end{align}
 	where $\bm{\Omega}_{\sf t}$ is a nested permutation matrix that changes the orders of transmit waveguides and segments on every waveguide, $\bm{\Omega}_{\sf r}$ is another nested permutation matrix that changes the orders of receive waveguides and segments on every waveguide, $\bm{\Pi}_{\sf t}, \bm{\Pi}_{\sf ue}, \bm{\Pi}_{\sf tar}$ are respectively permutation matrices that change the orders of transmit waveguides, users and targets. The detailed introduction of the nested permutation matrix can be found in \cite{guo2025attention}. The property indicates that the beamforming policy is not affected by the orders of transmit/receive waveguides, segments on every waveguide, users and targets.
 	
 	\section{SWISAC-AO: Optimizing Joint Beamforming with AO}\label{sec:optimization}
 	In this section, we propose the SWISAC-AO algorithm to optimize beamforming from problem \eqref{eq:prob}. The problem is non-convex, because the objective function in \eqref{eq:obj0} and the constraint in \eqref{eq:const-r} are non-convex to all the variables. To solve the problem efficiently, we first transform \eqref{eq:obj0} and \eqref{eq:const-r} such that they are at least convex to some of the variables.
 	\subsection{Problem Transformation}
 	\subsubsection{Transforming \eqref{eq:obj0}}
 	The objective function in \eqref{eq:obj0} can be regarded as the CR of $N_{\sf r}$ equivalent users (EUs), where $N_{\sf t}$ data streams are transmitted to each EU by $MN_{\sf t}$ antennas, and there is only one single antenna deployed at the EU. There is no interference among EUs. The receiving signal of the $n$-th EU can be expressed as $y_n = \tilde{\mathbf{h}}_{{\sf s},n}^{\sf H}\mathbf{W}\mathbf{s}_n + n_n$, where $\mathbf{s}_n$ and $n_n$ are respectively the transmitted signal and the receiving noise for the $n$-th EU,
 	$\tilde{\mathbf{h}}_{{\sf s},n}\triangleq \sqrt{L\alpha}\tilde{\mathbf{h}}(\bm{\Phi}_{\sf t},\mathbf{V}_{\sf tar}, \bm{\phi}_{{\sf r},n})$.
 	The estimated signal is given by $\hat{\mathbf{s}}_n=\mathbf{v}_ny_n=\mathbf{v}_n\tilde{\mathbf{h}}_{{\sf s},n}^{\sf H}\mathbf{W}\mathbf{s}_n + \mathbf{v}_nn_n$. It is not hard to obtain that the mean squared error of the estimation is $\mathbf{E}_n^{\sf MSE}\triangleq{\mathbb E}\{(\hat{\mathbf{s}}_n-\mathbf{s}_n)(\hat{\mathbf{s}}_n-\mathbf{s}_n)^{\sf H}\}=(\mathbf{I}_{K_{\sf c}}-\mathbf{v}_n\tilde{\mathbf{h}}_{{\sf s},n}^{\sf H}\mathbf{W})(\mathbf{I}_{K_{\sf c}}-\mathbf{v}_n\tilde{\mathbf{h}}_{{\sf s},n}^{\sf H}\mathbf{W})^{\sf H}+\sigma_{\sf s}^2\mathbf{v}_n\mathbf{v}_n^{\sf H}$. According to \cite{WMMSE2011Shi}, the objective function in \eqref{eq:obj0} is equivalent to the following objective function,
 	\begin{align}
 		\min_{\bm{\Phi}_{\sf t}, \bm{\Phi}_{\sf r}, \mathbf{W},\mathbf{V},\mathbf{{\bm\Theta}}} &L(\bm{\Phi}_{\sf t}, \bm{\Phi}_{\sf r}, \mathbf{W},\mathbf{V},\mathbf{{\bm\Theta}}) \triangleq \notag\\
 		&\sum_{n=1}^{N_{\sf r}} \mathsf{Tr}({\bm\Theta}_n\mathbf{E}_n^{\sf MSE})-\log\det({\bm\Theta}_n),
 	\end{align}
 	where ${\bm\Theta}_n$ is a weight matrix for the receiver of the $n$-th EU, $\bm{\Theta}=[{\bm\Theta}_1,\cdots,{\bm\Theta}_{N_{\sf r}}]$, $\mathbf{V}=[\mathbf{v}_1,\cdots,\mathbf{v}_{N_{\sf r}}]$ is the receiving beamforming matrix. It is not hard to prove that when $\{\bm{\Phi}_{\sf t}, \bm{\Phi}_{\sf r}\}$ are fixed, the objective function is convex to every variable in $\{\mathbf{W},\mathbf{V},\mathbf{{\bm\Theta}}\}$ when fixing other variables.
 	
 	\subsubsection{Transforming \eqref{eq:const-r}}
 	We resort to the method in \cite{Precod_Opt_Structure} to extract the hidden convexity of the constraint in \eqref{eq:const-r} w.r.t. $\mathbf{W}$. Specifically, the constraint can be re-expressed as,
 	\begin{align}\label{eq:const-1}
 		\frac{1}{\gamma\sigma_{\sf c}^2} |\tilde{\mathbf{h}}_k^{\sf H}\mathbf{w}_k|^2\geq \frac{1}{\sigma_{\sf c}^2}\sum_{j=1,j\neq k}^{K_{\sf c}} |\tilde{\mathbf{h}}_k^{\sf H}\mathbf{w}_j|^2 + 1,
 	\end{align}
 	where $\gamma \triangleq 2^{R_{\rm min}}-1$, $\tilde{\mathbf{h}}_{k}\triangleq \mathbf{G}^{\sf H}(\mathbf{\Phi}_{\sf t})\mathbf{h}(\mathbf{u}_k, \mathbf{\Phi}_{\sf t})$. By leveraging the phase ambiguity to rotate the phase such that the inner product $\tilde{\mathbf{h}}_k^{\sf H}\mathbf{w}_k$ is real and positive, \eqref{eq:const-1} can be rewritten as,
 	\begin{align}\label{eq:const-2}
 		\frac{1}{\sqrt{\gamma\sigma_{\sf c}^2}} \mathcal{R}(\tilde{\mathbf{h}}_k^{\sf H}\mathbf{w}_k)\geq \sqrt{\frac{1}{\sigma_{\sf c}^2}\sum_{j=1,j\neq k}^{K_{\sf c}} |\tilde{\mathbf{h}}_k^{\sf H}\mathbf{w}_j|^2 + 1},
 	\end{align}
 	which is a second-order cone constraint that is convex to $\mathbf{W}$.
 	
 	With these transformations, problem \eqref{eq:prob} can be transformed to,
 	\begin{subequations}\label{eq:opt-tb2}
 		\begin{align}
 			\min_{\bm{\Phi}_{\sf t}, \bm{\Phi}_{\sf r}, \mathbf{W},\mathbf{V},\mathbf{{\bm\Theta}}} & L(\bm{\Phi}_{\sf t}, \bm{\Phi}_{\sf r},\mathbf{W},\mathbf{V},\mathbf{{\bm\Theta}}) \label{eq:obj-t} \\ 
 			{\rm s.t.} ~ & \eqref{eq:const-2}, \eqref{eq:const-tx}, \eqref{eq:const-rx}, \eqref{eq:p-cons-min-d}, \eqref{eq:p-cons-power}.
 		\end{align}
 	\end{subequations}
 	
 	\subsection{Solving Problem \eqref{eq:opt-tb2} with AO}
 	
 	Since problem \eqref{eq:opt-tb2} is convex with respect to some variables when fixing other variables, we consider optimizing the problem via AO.
 	\subsubsection{Updating $\bm{\Phi}_{\sf t}$ and $\bm{\Phi}_{\sf r}$}
 	When $\mathbf{W}, \mathbf{V}, \bm{\Theta}$ are fixed, problem \eqref{eq:opt-tb2} reduces to the following problem, 
 	\begin{subequations}\label{eq:opt-tb}
 		\begin{align}
 			\min_{\bm{\Phi}_{\sf t}, \bm{\Phi}_{\sf r}} & L(\bm{\Phi}_{\sf t}, \bm{\Phi}_{\sf r},\mathbf{W},\mathbf{V},\mathbf{{\bm\Theta}}) \label{eq:obj-t2} \\ 
 			{\rm s.t.} ~ & \eqref{eq:const-2}, \eqref{eq:const-tx}, \eqref{eq:const-rx}, \eqref{eq:p-cons-min-d}.
 		\end{align}
 	\end{subequations}
 	The objective function and the constraint \eqref{eq:const-2} are highly non-convex w.r.t. $\bm{\Phi}_{\sf t}, \bm{\Phi}_{\sf r}$. This is mainly due to the existence of the periodic function $e^{-jx}$ in the free-space and in-waveguide channels. Hence, it is hard to use gradient-based methods for finding good enough PA positions. Non-gradient-based methods can be leveraged to find the positions, such as one-dimensional search or particle swarm optimization. Despite that, we derive the equation for updating PA positions with the gradient descent method, to gain insight for designing a DNN to learn the beamforming policy later.
 	
 	To this end, we first transform problem \eqref{eq:opt-tb} to a constraint-free version, where the constraints in \eqref{eq:const-2} are satisfied by the penalty function method. The objective function of the constraint-free optimization problem can be expressed as,
 	\begin{align}\label{eq:obj-const-free}
 		&\tilde{L}(\bm{\Theta}_{\sf t},\bm{\Theta}_{\sf r},\mathbf{W},\mathbf{V},\mathbf{{\bm\Theta}}) =\notag\\ &\sum_{n=1}^{N_{\sf r}}\! \mathsf{Tr}(\mathbf{{\bm\Theta}}_n ((\mathbf{I}_{K_{\sf c}}\!-\!\mathbf{v}_n\tilde{\mathbf{h}}_{{\sf s},n}^{\sf H}\mathbf{W})(\mathbf{I}_{K_{\sf c}}\!-\!\mathbf{v}_n\tilde{\mathbf{h}}_{{\sf s},n}^{\sf H}\mathbf{W})^{\sf H}\!\!+\!\sigma_{\sf s}^2\mathbf{v}_n\mathbf{v}_n^{\sf H}))-\notag\\& \log\det({\bm\Theta}_n) \!-\! 
 		\sum_{k=1}^{K_{\sf c}}\mu_k \Big(\frac{1}{\gamma\sigma_{\sf c}^2} |\tilde{\mathbf{h}}_k^{\sf H}\mathbf{w}_k|^2\!-\!\frac{1}{\sigma_{\sf c}^2}\!\!\sum_{j=1,j\neq k}^{K_{\sf c}}\!\! |\tilde{\mathbf{h}}_k^{\sf H}\mathbf{w}_j|^2 \!-\! 1\Big),
 		% -  \sum_{m,n,s} \xi_{mns} \big(x_{{\sf t}, msn} - x_{{\sf t}, (m-1)sn} - \Delta_{\min}\big),
 	\end{align}
 	where $\mu_k\geq 0,\forall k$ are penalty factors. The constraints in \eqref{eq:const-rx}, \eqref{eq:const-tx}, \eqref{eq:p-cons-min-d} are not considered in \eqref{eq:obj-const-free}, as they can be satisfied by projecting the updated PA positions to the feasible region.
 	
 	The equation for updating the transmit PA positions with the gradient descent method can be expressed as,
 	\begin{align}\label{eq:upd-x}
 		x_{{\sf t}, msn} \!\leftarrow& x_{{\sf t}, msn}  \!-\! \psi \frac{\partial \tilde{L}(\bm{\Phi}_{\sf t},\bm{\Phi}_{\sf r},\mathbf{W},\mathbf{V},\mathbf{{\bm\Theta}})}{\partial x_{{\sf t}, msn} } \notag\\
 		=&x_{{\sf t}, msn} \!-\! \psi\Bigg(\!
 		\sum_{k=1}^{K_{\sf s}}\!\frac{\partial \tilde{L}(\bm{\Theta}_{\sf t},\bm{\Theta}_{\sf r},\mathbf{W},\mathbf{V},\mathbf{{\bm\Theta}})}{\partial \mathbf{h}(\mathbf{u}_{{\sf tar},k},\bm{\Phi}_t)}
 		\frac{\partial \mathbf{h}(\mathbf{u}_{{\sf tar},k},\bm{\Phi}_t)}{\partial x_{{\sf t}, msn}}  \notag\\&
 		+\frac{\partial \tilde{L}(\bm{\Theta}_{\sf t},\bm{\Theta}_{\sf r},\mathbf{W},\mathbf{V},\mathbf{{\bm\Theta}})}{\partial \mathbf{G}(\bm{\Phi}_t)}
 		\frac{\partial \mathbf{G}(\bm{\Phi}_t)}{\partial x_{{\sf t}, msn}}+\notag\\
 		&\sum_{k=1}^{K_{\sf c}}\frac{\partial \tilde{L}(\bm{\Theta}_{\sf t},\bm{\Theta}_{\sf r},\mathbf{W},\mathbf{V},\mathbf{{\bm\Theta}})}{\partial \mathbf{h}(\mathbf{u}_k,\bm{\Phi}_t)}
 		\frac{\partial \mathbf{h}(k,\bm{\Phi}_t)}{\partial x_{{\sf t}, msn}}\Bigg),
 	\end{align} 
 	where the derivates of $\tilde{L}$ are given in \eqref{eq:gradient-1}$\sim$\eqref{eq:gradient-3} on the top of the next page,
 	\begin{figure*}
 		\begin{align}
 			\frac{\partial \tilde{L}(\bm{\Theta}_{\sf t},\bm{\Theta}_{\sf r},\mathbf{W},\mathbf{V},\mathbf{{\bm\Theta}})}{\partial \mathbf{h}(\mathbf{u}_{{\sf tar},k},\bm{\Phi}_t)} &= -\sqrt{L\alpha}\sum_{n=1}^{N_{\sf r}} \Bigg(\sum_{s=1}^{S} g(\bm{\phi}_{{\sf r},sn})h(\mathbf{u}_{{\sf tar},k},\bm{\phi}_{{\sf r},sn})\Bigg)
 			\mathbf{G}(\bm{\Phi}_{\sf t})\mathbf{W}
 			\mathbf{A}_n^{\sf H}
 			\bm{\Theta}_n\mathbf{v}_n\label{eq:gradient-1}\\
 			%\frac{\partial \mathbf{h}(\mathbf{u}_{{\sf tar},k},\bm{\Phi}_t)}{\partial x_{{\sf t}, msn}} &= \notag\\
 			\frac{\partial \tilde{L}(\bm{\Theta}_{\sf t},\bm{\Theta}_{\sf r},\mathbf{W},\mathbf{V},\mathbf{{\bm\Theta}})}{\partial \mathbf{G}(\bm{\Phi}_t)}  &= -\sum_{n=1}^{N_{\sf r}}
 			\mathbf{W}\mathbf{A}_n^{\sf H}\bm{\Theta}_n\mathbf{v}_n\mathbf{r}_n^{\sf H}
 			\;+\;
 			\frac{1}{\sigma_{\sf c}^2}\sum_{k=1}^{K_{\sf c}}\mu_k
 			\sum_{j=1}^{K_{\sf c}}\mathbf{h}(\mathbf{u}_k,\bm\Phi_{\sf t})(\mathbf{B})_{kj}\mathbf{w}_j^{\sf H}\label{eq:gradient-2}\\
 			%\frac{\partial \mathbf{G}(\bm{\Phi}_t)}{\partial x_{{\sf t}, msn}}&= \notag\\
 			\frac{\partial \tilde{L}(\bm{\Theta}_{\sf t},\bm{\Theta}_{\sf r},\mathbf{W},\mathbf{V},\mathbf{{\bm\Theta}})}{\partial \mathbf{h}(\mathbf{u}_k,\bm{\Phi}_t)} &= \frac{\mu_k}{\sigma_{\sf c}^2}\mathbf{G}(\bm{\Phi}_t)
 			\sum_{j=1}^K\mathbf{w}_j(\mathbf{B})_{kj}^*\label{eq:gradient-3}
 			%\frac{\partial \mathbf{h}(k,\bm{\Phi}_t)}{\partial x_{{\sf t}, msn}} = 
 		\end{align}
 	\end{figure*}
 	$\mathbf{r}_n \triangleq\sqrt{L\alpha}\sum_{k=1}^{K_{\sf s}}\sum_{s=1}^{S} $ $g(\bm{\phi}_{{\sf r},sn})h(\mathbf{u}_{{\sf tar},k},\bm{\phi}_{{\sf r},sn})\mathbf{h}(\mathbf{u}_{{\sf tar},k},\bm{\Phi}_{\sf t})$,
 	%$\tilde{\mathbf{h}}_{{\sf s},n}^{\sf H} \triangleq \mathbf{r}_n^{\sf H}\mathbf{G}(\bm\Phi_{\sf t})$,
 	$\mathbf{A}_n \triangleq \mathbf{I}_{K_{\sf c}}-\mathbf{v}_n\tilde{\mathbf{h}}_{{\sf s},n}^{\sf H}\mathbf{W}$, 
 	$\mathbf{B}$ is a $K_{\sf c}\times K_{\sf c}$ matrix with $(\mathbf{B})_{kj}$ being the element on its $k$-th row and $j$-th column. $(\mathbf{B})_{kj}=\mathbf{h}^{\sf H}(\mathbf{u}_k,\bm\Phi_{\sf t})\mathbf{G}(\bm\Phi_{\sf t})\mathbf{w}_j$ if $k\neq j$ and $(\mathbf{B})_{kj}=-\frac{1}{\gamma}\mathbf{h}^{\sf H}(\mathbf{u}_k,\bm\Phi_{\sf t})\mathbf{G}(\bm\Phi_{\sf t})\mathbf{w}_j$ if $k=j$. 
 	$\frac{\partial \mathbf{h}(\mathbf{u}_{{\sf tar},k},\bm{\Phi}_t)}{\partial x_{{\sf t}, msn}},\frac{\partial \mathbf{G}(\bm{\Phi}_t)}{\partial x_{{\sf t}, msn}},\frac{\partial \mathbf{h}(k,\bm{\Phi}_t)}{\partial x_{{\sf t}, msn}}$ can be derived from \eqref{eq:chl} and \eqref{eq:in-wg-chl} and hence are not provided here.

 	\subsubsection{Updating $\mathbf{V}$}
 	By solving $\frac{\partial L(\bm{\Phi}_{\sf t}, \bm{\Phi}_{\sf r},\mathbf{W, V},\bm{\Theta})}{\partial \mathbf{v}_n}=\mathbf{0}$, it is not hard to obtain the updated value of $\mathbf{v}_n$ as,
 	\begin{align}\label{eq:upd-v}
 		\mathbf{v}_n \leftarrow \frac{(\mathbf{W})^{\sf H}\tilde{\mathbf{h}}_{{\sf s},n}}{\|\tilde{\mathbf{h}}_{{\sf s},n}^{\sf H}\mathbf{W}\|^2+\sigma_{\sf s}^2}.
 	\end{align} 
 	
 	\subsubsection{Updating $\bm{\Theta}$}
 	By solving $\frac{\partial L(\bm{\Phi}_{\sf t}, \bm{\Phi}_{\sf r},\mathbf{W, V},\bm{\Theta})}{\partial \bm{\Theta}_n}=\mathbf{0}$, it is not hard to obtain the updated value of $\bm{\Theta}$  as,
 	\begin{align}\label{eq:opt-theta}
 		\bm{\Theta}_n \leftarrow (\mathbf{E}_n^{{\sf MSE}})^{-1}.
 	\end{align}
 	% where $\mathbf{E}_n^{{\sf MSE}{(\ell)}}=(\mathbf{I}_{K_{\sf c}}-\mathbf{v}_n^{(\ell)}\tilde{\mathbf{h}}_{{\sf s},n}^{\sf H}\mathbf{W}^{(\ell)})(\mathbf{I}_{K_{\sf c}}-\mathbf{v}_n^{(\ell)}\tilde{\mathbf{h}}_{{\sf s},n}^{\sf H}\mathbf{W}^{(\ell)})^{\sf H}+\sigma_{\sf s}^2\mathbf{v}_n^{(\ell)}(\mathbf{v}_n^{(\ell)})^{\sf H}$.
 	
 	\subsubsection{Updating $\mathbf{W}$}
 	When other variables are fixed, $\mathbf{W}$ can be solved from the following problem,
 	\begin{subequations}\label{eq:opt-tb3}
 		\begin{align}
 			\min_{\mathbf{W}}~ & L(\bm{\Phi}_{\sf t}, \bm{\Phi}_{\sf r},\mathbf{W},\mathbf{V},\mathbf{{\bm\Theta}}) \label{eq:obj-t1} \\ 
 			{\rm s.t.} ~ & \eqref{eq:const-2},\eqref{eq:p-cons-power}
 		\end{align}
 	\end{subequations}
 	
 	The problem is convex and can be solved by resorting to convex optimization tools such as CVX.

 	To gain useful insight into the GNN design to be introduced later, we derive an optimal solution structure of the problem.
 	\begin{proposition}\label{prop:structure}
 		The optimal transmit beamforming is with the structure of \eqref{eq:structure} in the next page, where $\sum_{k=1}^{K_{\sf c}}p_k=P_{\max}$, $\xi_{nk}\triangleq (\bm{\Theta}_n\mathbf{v}_n)_k/(\frac{(1+\gamma)\beta_k}{\gamma\sigma_{\sf c}^2}(\tilde{\bf h}_k^{\sf H}\mathbf{w}_k))$, $\nu\geq0, \beta_k\geq 0,k=1,\cdots,K_{\sf c}$ are the Lagrange multipliers. 
 		\begin{IEEEproof}
 			See Appendix \ref{appendix:structure}.
 		\end{IEEEproof}
 	\end{proposition}   
 	\textbf{Observation 1}: The transmit beamforming vector lies in the subspace spanned by $\tilde{\mathbf{h}}_k, \tilde{\mathbf{h}}_{{\sf s},n}, n=1,\cdots,N_{\sf r}$.
 	\begin{figure*}
 		\begin{align}\label{eq:structure}
 			\mathbf{w}_k^{\star} = \sqrt{p_k}\frac{\Big(\sum_{n=1}^{N_{\sf r}} (\mathbf{v}_n^{\sf H}\bm{\Theta}_n\mathbf{v}_n)\tilde{\mathbf{h}}_{{\sf s},n}\tilde{\mathbf{h}}_{{\sf s},n}^{\sf H}+\sum_{j=1}^{K_{\sf c}} \frac{\beta_j}{\sigma_{\sf c}^2}\tilde{\mathbf{h}}_j\tilde{\bf h}_j^{\sf H}+\nu\mathbf{I}\Big)^{-1}\Big(\tilde{\mathbf{h}}_k+\sum_{n=1}^{N_{\sf r}}\xi_{nk}\tilde{\bf h}_{{\sf s},n}\Big)}{\Big\|\Big(\sum_{n=1}^{N_{\sf r}} (\mathbf{v}_n^{\sf H}\bm{\Theta}_n\mathbf{v}_n)\tilde{\mathbf{h}}_{{\sf s},n}\tilde{\mathbf{h}}_{{\sf s},n}^{\sf H}+\sum_{j=1}^{K_{\sf c}} \frac{\beta_j}{\sigma_{\sf c}^2}\tilde{\mathbf{h}}_j\tilde{\bf h}_j^{\sf H}+\nu\mathbf{I}\Big)^{-1}\Big(\tilde{\mathbf{h}}_k+\sum_{n=1}^{N_{\sf r}}\xi_{nk}\tilde{\bf h}_{{\sf s},n}\Big)\Big\|}
 		\end{align}
 	\end{figure*}
 	where $\sum_{k=1}^{K_{\sf c}}p_k=P_{\max}$, $\xi_{nk}\triangleq (\bm{\Theta}_n\mathbf{v}_n)_k/(\frac{(1+\gamma)\beta_k}{\gamma\sigma_{\sf c}^2}(\tilde{\bf h}_k^{\sf H}\mathbf{w}_k))$.
 	
 	The SWISAC-AO algorithm is summarized in Algorithm \ref{algo:opt}.
 	\begin{algorithm}
 		\caption{SWISAC-AO algorithm for solving problem \eqref{eq:prob}}\label{algo:opt}
 		\small
 		\begin{algorithmic}[1]
 			\State \textbf{Input}: $\mathbf{U},\mathbf{U}_{\sf tar}$, Number of iterations $T$
 			\State \textbf{Output}: $\bm\Phi_{\sf t},\bm\Phi_{\sf r}, \mathbf{W}$.
 			\State Initialize $\bm\Phi_{\sf t}, \bm\Phi_{\sf r}, \mathbf{W}, \mathbf{V}, \bm\Theta$
 			\For {$t=1:T$}
 			\State Update $x_{{\sf t},msn}$ and $x_{{\sf r},ms}$ with one-dimensional search, $\forall m,s,n$
 			\State $\bm\phi_{{\sf t},msn}=[x_{{\sf t},msn}, y_{{\sf t},n}, z_{{\sf t},n}]$, $\bm\phi_{{\sf r},ms}=[x_{{\sf r},ms}, y_{{\sf r},n}, z_{{\sf r},n}]$.
 			\State Update $\mathbf{W}$ by solving problem \eqref{eq:opt-tb3} with CVX.
 			\State Update $\mathbf{v}_n$ with \eqref{eq:upd-v}, $\forall n$.
 			\State Update $\bm\Theta_n$ with \eqref{eq:opt-theta}, $\forall n$.
 			\EndFor

 		\end{algorithmic}
 	\end{algorithm}

	\section{SWISAC-GNN: Learning Joint Beamforming with GNN} \label{sec:learn-bf}
	The SWISAC-AO algorithm can solve problem \eqref{eq:prob} but can be computationally intensive because a sub-problem needs to be solved in each iteration. To enable real-time implementation, we propose a GNN architecture to learn beamforming. 
	%For notational simplicity, we consider $M=1$ in the sequel, i.e., there is only one PA on each waveguide segment.
	
	% \subsection{Predicting Positions of Users and Targets}
	% As Transformers employ the attention mechanism that can effectively model temporal correlation \cite{Attn}, it can be used to predict the positions of users and targets. 
	% Since the architecture of Transformer has been well-introduced in \cite{Attn}, it is not introduced in detail for the conciseness of the paper.
	% The Transformer is shared among all the users and targets.
	
	% \subsection{Learning Beamforming with the Predicted Positions}
	\subsection{Graph Modeling}
\begin{figure*}[!htb]
		\centering
		\includegraphics[width=.7\linewidth]{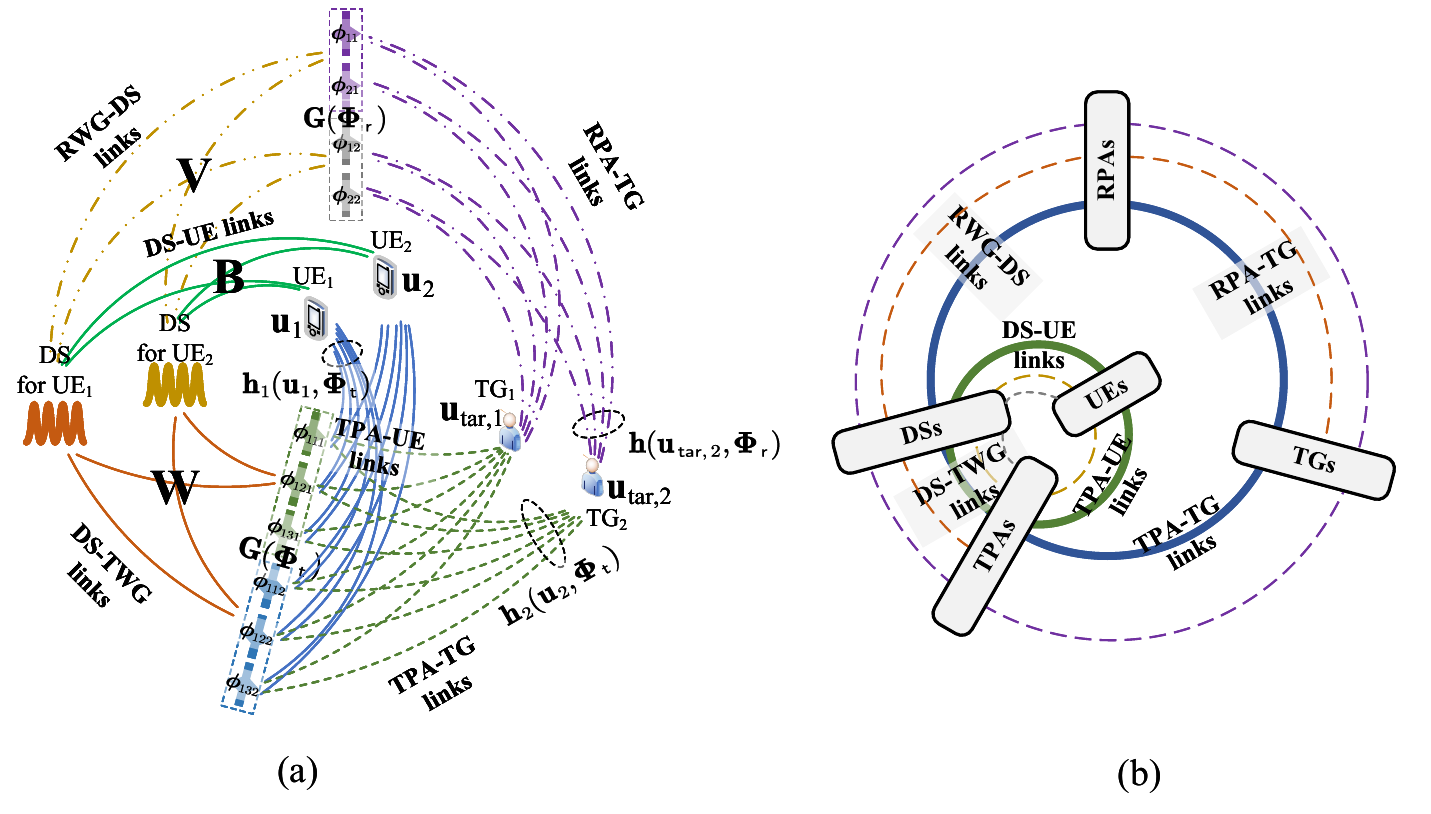}
		\caption{Illustration of modeled graph, $K=2, N=3$.}
		\label{fig:graph}
	\end{figure*}

	The PASS-assisted ISAC system can be modeled as a graph as in Fig. \ref{fig:graph}. 
	The vertices and edges may be associated with features and actions, which are detailed as follows.
	
	\begin{itemize}
		\item \textbf{Vertices}: There are five types of vertices in the graph.
		\begin{itemize}
			\item Data streams (DSs): There are no features or actions on DSs.
			\item Transmitting PAs (TPAs): There is no feature on TPAs. The action on the $m$-th TPA of the $s$-th segment of the $n$-th transmitting waveguide (TWG) is $\bm{\phi}_{{\sf t}, msn}$.
			\item Receiving PA (RPA): There is no feature on the RPA. The action on the RPA of the $s$-th segment of the $n$-th receiving waveguide (RWG) is $\bm{\phi}_{{\sf r},sn}$.
			\item Users (UEs): The feature on the $k$-th UE is $\mathbf{u}_k$. There is no action on each UE. 
			\item Targets (TGs): The feature on the $k$-th TG is $\mathbf{u}_{{\sf tar}, k}$. There is no action on each TG. 
		\end{itemize}
		\item \textbf{Edges}: There are six types of edges in the graph, which are respectively the DS-TWG links, TPA-UE links, TPA-TG links, RPA-TG links, RWG-DS links and DS-UE links. The actions on the DS-TWG links form the transmit beamforming matrix $\mathbf{W}$. The actions on the RWG-DS links form the receiving beamforming matrix $\mathbf{V}$.
	\end{itemize}
	
	In addition to features and actions, the channel coefficients (including both the in-waveguide and the free-space channels) and MUI can be computed from the features and actions with mathematical models. They are called \emph{representations} of the edges or vertices, which are listed below.
	\begin{itemize}
		\item TPAs:
		The representations of TPAs are the in-waveguide channel coefficients. For example, the representations of the TPAs on the $s$-th segment of the $n$-th TWG form a vector $\mathbf{g}(\bm{\Phi}_{{\sf t},sn})$, whose expression is provided in \eqref{eq:in-wg-chl}. The representations of all TPAs form $\mathbf{G}(\bm{\Phi}_{\sf t})$.
		\item RPAs:
		The representations of RPAs are also the in-waveguide channel coefficients. For example, for the RPA on the $s$-th segment of the $n$-th RWG, the representation is $g(\bm{\Phi}_{{\sf r},sn})$. The representations of all TPAs form $\mathbf{G}(\bm{\Phi}_{\sf r})$.
		\item TPA-UE links:
		The representations of the TPA-UE links are the free-space channel coefficients between them. For example, for the link from the $m$-th TPA on the $s$-th segment of the $n$-th TWG (called TPA$_{msn}$) to the $k$-th user, the representation is $h(\mathbf{u}_k,\bm{\phi}_{{\sf t}, msn})$. 
		\item TPA-TG links:
		The representations of the TPA-TG links are the free-space channel coefficients between them. For example, for the link from TPA$_{msn}$ to the $k$-th target, the representation is $h(\mathbf{u}_{{\sf tar},k},\bm{\phi}_{{\sf t}, msn})$. 
		\item RPA-TG links:
		The representations of the RPA-TG links are the free-space channel coefficients between them. For example, for the link from the $k$-th target to the RPA on the $s$-th segment of the $n$-th RWG, the representation is $h(\mathbf{u}_{{\sf tar},k},\bm{\phi}_{{\sf r}, sn})$. 
		\item DS-UE links: The representations of the DS-UE links form the matrix $\mathbf{B}$, which reflects MUI.
	\end{itemize}
	
	To see how the vertices and edges of different types are connected more clearly, we provide a simplified version of the modeled graph in Fig. \ref{fig:graph}(b). Each box shows one type of vertices and the solid curves among them show the edges. We can see that there are two loops in the graph, respectively shown by the blue and green solid circles.
	
	\subsection{GNN Design}
	The beamforming policy can be learned over the graph by learning the mapping from features to actions. There are many ways of designing GNNs for learning over the graph, but not all of them perform well. Without judicious design, information loss or permutation property mismatch issues may occur, which degrade the learning performance \cite{LSJ}.
	
	The analogy between the iterative equation of a numerical algorithm and the equation of updating the actions of GNNs (called the update equation) has been noticed in \cite{GJ-RGNN}. Specifically, the numerical algorithm approximates a policy (say beamforming) with multiple iterations, which is analogous to a GNN that learns the policy after multiple layers. It was proved in \cite{GJ-RGNN} that the iterative equation of an algorithm can be re-expressed as the update equation of a GNN. 
	
	Based on the analogy, we design the GNN by drawing inspiration from the iterative equation of the SWISAC-AO algorithm. Since the algorithm alternatively updates the PA positions and the transmit/receiving beamforming matrices in each iteration, we design the GNN that also alternatively updates them, i.e., the actions on the TPAs, RPAs, DS-TWG links and RWG-DS links. 
	
	For updating the action of a vertex or an edge in the $(\ell+1)$-th layer, the update equation of a GNN typically involves two components,
	\begin{itemize}
		\item \emph{Aggregation}: The actions or representations of the vertices and edges neighboring to the vertex or the edge are processed by a \emph{processor} for extracting information from the actions or representations. The processor is a parameterized function, e.g., FNN. Then, the extracted information is aggregated by a \emph{pooling function}, say summation.
		\item \emph{Combination}: The aggregated information is combined with the actions or representations of the vertex or edge itself in the $\ell$-th layer to update its action in the $(\ell+1)$-th layer. The combiner is also a parameterized function.
	\end{itemize}
	
	Since the graph in Fig. \ref{fig:graph} is a heterogeneous graph with multiple types of vertices and edges, the information aggregation from other types of vertices and edges is involved in the update equation. The design of information aggregation can be inspired by the iterative equation of the SWISAC-AO algorithm. To see this more clearly, we omit some update variables that less affect the performance of the algorithm by making the following assumptions.
	\begin{enumerate}
		\item The EUs can recover the transmitted signal with almost no error, i.e., $\mathbf{v}_n\tilde{\mathbf{h}}_{{\sf s},n}^{\sf H}\mathbf{W}\approx \eta\mathbf{I}_{K_{\sf c}}$, where $\eta$ is a factor that reflects the strength of the received signal.
		\item The noise power at the RPAs is very low, i.e., $\sigma_{\sf s}^2\approx 0$.
	\end{enumerate}
	
	With the two assumptions, and \eqref{eq:opt-theta}, it is not hard to obtain that $\mathbf{A}_n\bm{\Theta}_n\approx \frac{1}{1-\eta}\mathbf{I}_{K_{\sf c}}$.
	
	\subsubsection{Update the PA positions}
	Under the assumption, the partial derivatives \eqref{eq:gradient-1}$\sim$\eqref{eq:gradient-3} can be approximated as \eqref{eq:gradient-1-1}$\sim$\eqref{eq:gradient-3-1} on the top of the next page.
	\begin{figure*}
		\begin{align}
			\frac{\partial \tilde{L}(\bm{\Theta}_{\sf t},\bm{\Theta}_{\sf r},\mathbf{W},\mathbf{V},\mathbf{{\bm\Theta}})}{\partial \mathbf{h}(\mathbf{u}_{{\sf tar},k},\bm{\Phi}_t)} &\approx -\frac{\sqrt{L\alpha}}{1-\eta}\sum_{n=1}^{N_{\sf r}} \Bigg(\sum_{s=1}^{S} g(\bm{\phi}_{{\sf r},sn})h(\mathbf{u}_{{\sf tar},k},\bm{\phi}_{{\sf r},sn})\Bigg)
			\mathbf{G}(\bm{\Phi}_{\sf t})\mathbf{W}
			\mathbf{v}_n\triangleq \nabla\tilde{L}_{{\bf h}_{\sf tar},k}(\bm\Phi_{\sf t},\bm\Phi_{\sf r},\mathbf{W},\mathbf{V})\label{eq:gradient-1-1}\\
			%\frac{\partial \mathbf{h}(\mathbf{u}_{{\sf tar},k},\bm{\Phi}_t)}{\partial x_{{\sf t}, msn}} &= \notag\\
			\frac{\partial \tilde{L}(\bm{\Theta}_{\sf t},\bm{\Theta}_{\sf r},\mathbf{W},\mathbf{V},\mathbf{{\bm\Theta}})}{\partial \mathbf{G}(\bm{\Phi}_t)}  &\approx -\frac{1}{1-\eta}\sum_{n=1}^{N_{\sf r}}
			\mathbf{W}\mathbf{v}_n\mathbf{r}_n^{\sf H}
			\;+\;
			\frac{1}{\sigma_{\sf c}^2}\sum_{k=1}^{K_{\sf c}}\beta_k
			\sum_{j=1}^{K_{\sf c}}\mathbf{h}(\mathbf{u}_k,\bm\Phi_{\sf t})(\mathbf{B})_{kj}\mathbf{w}_j^{\sf H}\triangleq \nabla\tilde{L}_{\bf G}(\bm\Phi_{\sf t},\bm\Phi_{\sf r},\mathbf{W},\mathbf{V})
			\label{eq:gradient-2-1}\\
			%\frac{\partial \mathbf{G}(\bm{\Phi}_t)}{\partial x_{{\sf t}, msn}}&= \notag\\
			\frac{\partial \tilde{L}(\bm{\Theta}_{\sf t},\bm{\Theta}_{\sf r},\mathbf{W},\mathbf{V},\mathbf{{\bm\Theta}})}{\partial \mathbf{h}(\mathbf{u}_k,\bm{\Phi}_t)} &= \frac{\beta_k}{\sigma_{\sf c}^2}\mathbf{G}(\bm{\Phi}_t)
			\sum_{j=1}^K\mathbf{w}_j(\mathbf{B})_{kj}^*\triangleq \nabla\tilde{L}_{{\bf h},k}(\bm\Phi_{\sf t},\bm\Phi_{\sf r},\mathbf{W},\mathbf{V}),\label{eq:gradient-3-1}
			%\frac{\partial \mathbf{h}(k,\bm{\Phi}_t)}{\partial x_{{\sf t}, msn}} = 
		\end{align}
	\end{figure*}
	
	The three derivatives can be seen as the gradients for updating the representations of TPA-TG links, TWGs and TPA-UE links, respectively, where we can see how the information of other types of vertices and edges is aggregated with matrix multiplication. To see this more clearly,
	\setlength{\leftmargini}{1.2em}
	\begin{itemize}
		\item In \eqref{eq:gradient-1-1}, noticing that $g(\bm\phi_{{\sf r}, sn}), h(\mathbf{u}_{{\sf tar},k},\bm\phi_{{\sf r},sn}), \mathbf{G}(\bm\Phi_{\sf t})$, $ \mathbf{W}, \mathbf{v}_n$ are respectively the representations of RPAs, RPA-TG links, TPAs, DS-TWG links and RWG-DS links, the two summations and the matrix multiplications aggregate the information from these vertices and edges to TPA-TG links, in an end-to-end manner. To understand this more straightforwardly, we can see from Fig. \ref{fig:graph}(b) that there exists a ``hyper-link'' connected by RPA-TG links -- RPAs -- RWG-DS links -- DS-TWG links (shown by orange dashed curves). The information of these links is aggregated by summations and matrix multiplications.
		\item In \eqref{eq:gradient-2-1}, we can see that the two terms reflect information aggregation to the TPA vertices via two ``hyper-links''. The first hyper-link is created by TPA-TG links -- RPA-TG links -- RPAs -- DS-RWG links -- DS-TWG links. The second hyper-link is created by DS-TWG links -- DS-UE links -- TPA-UE links. The two hyper-links are respectively shown by purple and yellow dashed circles in Fig. \ref{fig:graph}(b).
		\item In \eqref{eq:gradient-3-1}, we can see that the matrix multiplication and summation reflect information aggregation from a ``hyper-link'' created by TPAs -- DS-TWG links -- DS-UE links. The hyper-link is shown by gray dashed curves in Fig. \ref{fig:graph}(b).
	\end{itemize}
	
	% From the aggregation process observed above, we can summarize how information from other types of vertices or edges should be aggregated.
	
	% \textbf{Observation 2}: To update the representations of one type of vertices or edges, the information of other types of vertices or edges should be aggregated via hyper-link(s) that can construct loop(s) with the vertices/edges itself.
	
	From \eqref{eq:upd-x}, we can see that with chain rules of derivatives, i.e., respectively multiplying the derivatives in \eqref{eq:gradient-1-1}$\sim$\eqref{eq:gradient-3-1} with $\frac{\partial \mathbf{h}(\mathbf{u}_{{\sf tar},k},\bm{\Phi}_t)}{\partial x_{{\sf t}, msn}},\frac{\partial \mathbf{h}(\mathbf{u}_{{\sf tar},k},\bm{\Phi}_t)}{\partial x_{{\sf t}, msn}}, \frac{\partial \mathbf{h}(k,\bm{\Phi}_t)}{\partial x_{{\sf t}, msn}}$, the information aggregated to TPA-TG links, TPAs, TPA-UE links can be used to update TPA positions. 
	
	Based on the observations above, we can design the equation for updating PA positions, where the information aggregation is inspired by the iterative equation. We take the update equation for TPA positions as an example. Denote the updated TPA, RPA positions, the transmit beamforming and combining matrices in the $\ell$-th layer as $\bm\Phi_{\sf t}^{(\ell)}=[\bm\phi_{{\sf t},111}^{(\ell)},\cdots,\bm\phi_{{\sf t},MSN}^{(\ell)}]$, $\bm\Phi_{\sf r}^{(\ell)}$, $\mathbf{W}^{(\ell)}$ and $\mathbf{V}^{(\ell)}$, respectively, where $\bm\phi_{{\sf t},MSN}^{(\ell)}=[x_{{\sf t},msn}^{(\ell)},y_{{\sf t},n},z_{{\sf t},n}]$. Inspired by the information aggregation observed above, the equation for updating $x_{{\sf t},msn}^{(\ell+1)}$ can be designed as,
	\begin{align}\label{eq:upd-tpa}
		&x_{{\sf t},msn}^{(\ell+1)} = \mathsf{FNN}\Bigg(x_{{\sf t},msn}^{(\ell)},\notag\\& \sum_{k=1}^{K_{\sf s}}
		\nabla\tilde{L}_{{\bf h}_{\sf tar},k}(\bm\Phi_{\sf t}^{(\ell)},\bm\Phi_{\sf r}^{(\ell)},\mathbf{W}^{(\ell)},\mathbf{V}^{(\ell)})
		\frac{\partial \mathbf{h}(\mathbf{u}_{{\sf tar},k},\bm{\Phi}_{\sf t}^{(\ell)})}{\partial x_{{\sf t}, msn}^{(\ell)}},\notag\\
		&\nabla\tilde{L}_{\bf G}(\bm\Phi_{\sf t}^{(\ell)},\bm\Phi_{\sf r}^{(\ell)},\mathbf{W}^{(\ell)},\mathbf{V}^{(\ell)})      
		\frac{\partial \mathbf{G}(\bm{\Phi}_{\sf t}^{(\ell)})}{\partial x_{{\sf t}, msn}^{(\ell)}},\notag\\
		&
		\sum_{k=1}^{K_{\sf c}}
		\nabla\tilde{L}_{{\bf h},k}(\bm\Phi_{\sf t}^{(\ell)},\!\bm\Phi_{\sf r}^{(\ell)},\!\mathbf{W}^{(\ell)},\!\mathbf{V}^{(\ell)})  
		\frac{\partial \mathbf{h}(k,\bm{\Phi}_{\sf t}^{(\ell)})}{\partial x_{{\sf t}, msn}^{(\ell)}}\Bigg),
	\end{align}
	where ${\sf FNN}(\cdot)$ denotes an FNN. By comparing \eqref{eq:upd-tpa} and \eqref{eq:upd-x}, we can find that the two equations are similar, and the differences are (i) in \eqref{eq:upd-tpa}, the derivatives of $\tilde{L}$ w.r.t. $\mathbf{h}(\mathbf{u}_{{\sf tar},k},\bm\Phi_{\sf t}^{(\ell)}),\mathbf{G}(\bm{\Phi}_{\sf t}^{(\ell)}),\mathbf{h}(\mathbf{u}_k,\bm{\Phi}_{\sf t}^{(\ell)})$ are replaced by the approximated values in \eqref{eq:gradient-1-1}$\sim$\eqref{eq:gradient-3-1}, and (ii) in \eqref{eq:upd-tpa}, a FNN is used to learn the information combination instead of using linear combination as in  \eqref{eq:upd-x}. This can improve the expression ability of GNN.
	
	The equation for updating the RPA positions can be designed similarly, hence is not provided here for the conciseness of the paper.
	
	\subsubsection{Update $\mathbf{W}$}
	To simplify the mapping to be learned by the GNN, we resort to the optimal solution structure in \eqref{eq:structure}. Specifically, we first update the unknown variables in \eqref{eq:structure}, including $\mathbf{p}=[p_1,\cdots,p_{K_{\sf c}}]$, $\bm\beta=[\beta_1,\cdots,\beta_{K_{\sf c}}]$ and $\bm\Xi$ ($(\bm\Xi)_{nk}=\xi_{nk}$)\footnote{We found that the value of $\nu$ in \eqref{eq:structure} has marginal effect on the learning performance, because the Lagrange multiplier is used to satisfy the power constraint \eqref{eq:p-cons-power}, which is already satisfied by power normalization. Hence, we set $\nu=1$.}, then, we recover the beamforming vectors with \eqref{eq:structure}. 
	
	$\mathbf{p}$ and $\bm\beta$ can be seen as the intermediate actions on the UE vertices, while $\bm\Xi$ can be seen as the intermediate actions on the edges between UE vertices and RPA vertices. To update them in the $(\ell+1)$-th layer (respectively denoted as $\bm\beta^{(\ell+1)}=[\beta_1^{(\ell+1)},\cdots,\beta_{K_{\sf c}}^{(\ell+1)}]$ and $\bm\Xi^{(\ell+1)}$), the information aggregation can be designed by matrix multiplication, while the information combination can be again designed as a FNN. 
	
	Taking the update of $\beta_{k}^{(\ell+1)}$ as an example, the information of vertices and edges can be aggregated via the hyper-link that is constructed by TPA-UE links -- TPAs -- DS-TWG links -- DS-UE links. Further considering the representations on these vertices/edges, the update equation can be designed as,
	\begin{align}\label{eq:upd-beta}
		\beta_{k}^{(\ell+1)} = \mathsf{FNN}\Big(\beta_{k}^{(\ell)}, \mathbf{h}^{\sf H}(\mathbf{u}_k, \bm\Phi_{\sf t}^{(\ell)})\mathbf{G}(\bm\Phi_{\sf t}^{(\ell)})\mathbf{W}^{(\ell)}\mathbf{b}_k^{(\ell)}\Big),
	\end{align}
	where $\mathbf{b}_k^{(\ell)}$ is the $k$-th column of $\mathbf{B}^{(\ell)}$, which is the updated value of $\mathbf{B}$ in the $\ell$-th layer.
	
	\subsubsection{Update $\mathbf{V}$} 
	To design the equation of updating $\mathbf{V}^{(\ell+1)}$ (i.e., the actions of RWG-DS links), the information combination can be again designed as a FNN, and the information aggregation can be designed by drawing inspiration from the iterative equation in \eqref{eq:upd-v}. Specifically, by substituting the expression of $\tilde{\mathbf{h}}_{{\sf s},n}$ into \eqref{eq:upd-v}, it is not hard to find that the information is aggregated by matrix multiplication via a hyper-link created by DS-TWG links -- TPAs -- TPA-TG links -- RPA-TG links -- RPAs. From this, the update equation can be designed as,
	\begin{align}\label{eq:upd-v-gnn}
		\mathbf{v}_n^{(\ell+1)} = &\mathsf{FNN}\Big(\mathbf{v}_n^{(\ell)}, \sum_{k=1}^{K_{\sf c}}\mathbf{W}^{(\ell)}\mathbf{G}(\bm\Phi_{\sf t}^{(\ell)})\mathbf{h}(\mathbf{u}_{{\sf tar},k},\bm\Phi_{\sf t}^{(\ell)})\cdot\notag\\&\mathbf{h}^{\sf H}(\mathbf{u}_{{\sf tar},k},\bm\Phi_{\sf r}^{(\ell)})\mathbf{G}(\bm\Phi_{\sf r}^{(\ell)})\Big).
	\end{align}
	
	% By substituting the expression of $\tilde{\mathbf{h}}_{{\sf s},n}$ into \eqref{eq:upd-v} and by comparing \eqref{eq:upd-v-gnn} with \eqref{eq:upd-v}, it is not hard to see that the aggregation process in \eqref{eq:upd-v-gnn} is exactly the same with the matrix multiplication in \eqref{eq:upd-v}. This validates the correctness of Observation 1 for designing information aggregation.
	
	\begin{figure}[!htb]
		\centering
		\includegraphics[width=\linewidth]{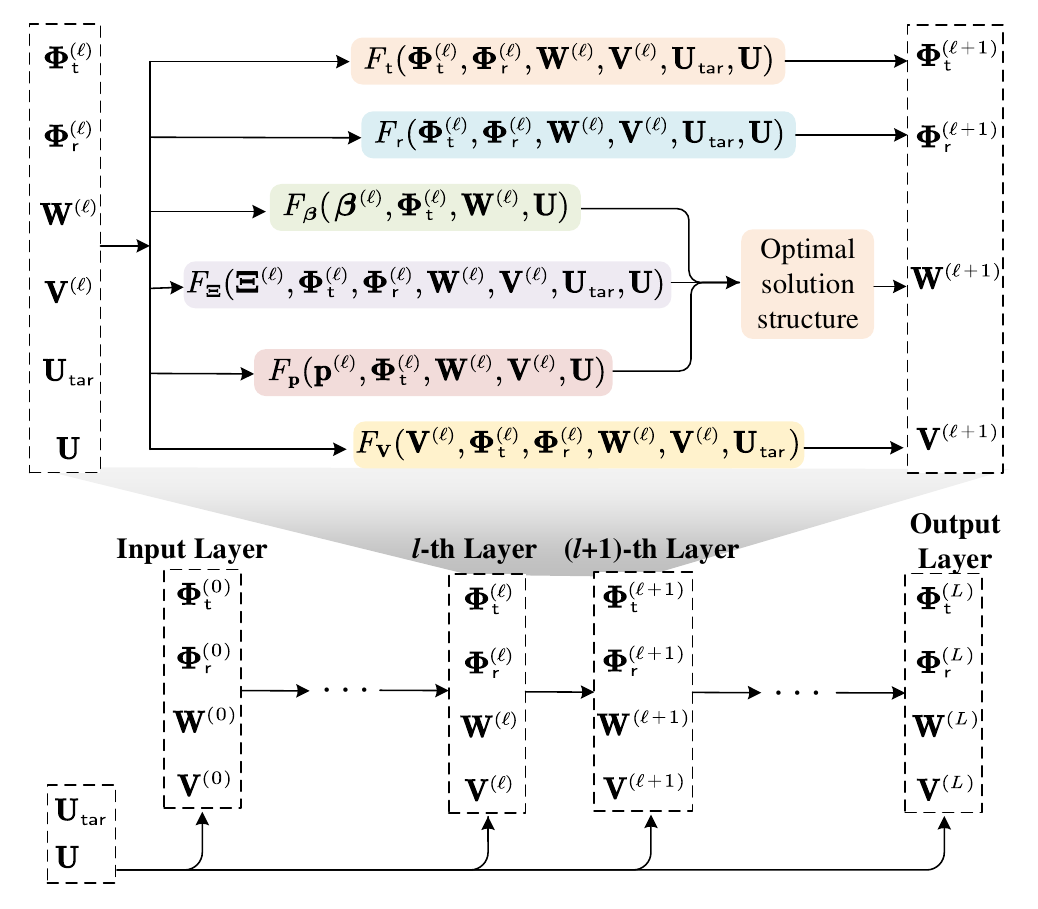}
		\caption{Architecture of SWISAC-GNN.}
		\label{fig:gnn}
	\end{figure}
	
	The GNN architecture is illustrated in Fig. \ref{fig:gnn}. In the figure, $F_{\sf t}(\cdot), F_{\sf r}(\cdot), F_{\bm\beta}(\cdot), F_{\bm\Xi}(\cdot), F_{\mathbf{p}}(\cdot), F_{\mathbf{V}}(\cdot)$ respectively denotes the equation for updating $\bm\Phi_{\sf t}^{(\ell+1)},\bm\Phi_{\sf r}^{(\ell+1)}$, $\bm\beta^{(\ell+1)},\bm\Xi^{(\ell+1)},\mathbf{p}^{(\ell+1)},\mathbf{V}^{(\ell+1)}$. For example, from \eqref{eq:upd-beta}, $F_{\bm\beta}(\cdot)$ can be expressed as,
	\begin{align}
		\bm\beta^{(\ell+1)} &= F_{\bm\beta}(\bm\beta^{(\ell)},\mathbf{W}^{(\ell)},\bm\Phi_{\sf t}^{(\ell)},\mathbf{U}) \notag\\
		&=\begin{bmatrix}
			\mathsf{FNN}\Big(\beta_{1}^{(\ell)}, \mathbf{h}^{\sf H}(\mathbf{u}_1, \bm\Phi_{\sf t}^{(\ell)})\mathbf{G}(\bm\Phi_{\sf t}^{(\ell)})\mathbf{W}^{(\ell)}\mathbf{b}_1^{(\ell)}\Big)\\
			\vdots\\
			\mathsf{FNN}\Big(\beta_{K}^{(\ell)}, \mathbf{h}^{\sf H}(\mathbf{u}_K, \bm\Phi_{\sf t}^{(\ell)})\mathbf{G}(\bm\Phi_{\sf t}^{(\ell)})\mathbf{W}^{(\ell)}\mathbf{b}_K^{(\ell)}\Big)
		\end{bmatrix}.\notag
	\end{align}
	
	\begin{proposition}\label{prop:pe}
		The input-output relation of the SWISAC-GNN can satisfy the permutation property in \eqref{eq:pe}.
		\begin{IEEEproof}
			See Appendix \ref{appendix}.
		\end{IEEEproof}
	\end{proposition}
	
	This proposition demonstrates that the beamforming policy lies in the hypothesis space of the GNN. This is important for learning the beamforming policy efficiently \cite{GJ_TWC_GNN}.
	
	\subsection{Training the GNN}
	To avoid generating labels that is time-consuming, the GNN is trained in an unsupervised manner. To maximize the objective function in \eqref{eq:obj0} while satisfying the constraints, the Lagrange multiplier method can be adopted \cite{mark2019learning}. Specifically, the loss function can be expressed as,
	\begin{align}
		\mathcal{L}=&-\frac{1}{N_{\sf tr}}\sum_{i=1}^{N_{\sf tr}} \Bigg(R_{\sf s}(\bm\Phi_{\sf t}^{[i]},\mathbf{U}_{\sf tar}^{[i]},\bm\Phi_{\sf r}^{[i]})+\notag\\
		&\sum_{k=1}^{K_{\sf c}}\beta_k(\mathbf{U}^{[i]},\mathbf{U}_{\sf tar}^{[i]})\big(R_{\rm min}-R_k(\mathbf{U}^{[i]},\bm\Phi_{\sf t}^{[i]})\big)\Bigg),\notag
	\end{align}
	where $N_{\sf tr}$ is the number of training samples and the subscript $[i]$ indicates the $i$-th sample. $\beta_k(\mathbf{U}^{[i]},\mathbf{U}_{\sf tar}^{[i]})\geq 0$ is the Lagrange multiplier for satisfying the constraint \eqref{eq:const-r}. It is a parameterized function that can be learned by a GNN. Other constraints in problem \eqref{eq:prob} are not considered in the loss function, because they can be satisfied by designing the activation functions in the output layer of GNN \cite{guo2025gpass}.

	\section{Numerical Results} \label{sec:simulation}
	In this section, we evaluate the performance of the SWISAC-GNN architecture and the SWISAC-AO algorithm with simulations.
	
	Consider a PASS-aided UAV system, where $N_{\sf t}$ TWGs and $N_{\sf r}$ RWGs are deployed on a building. 
	The TWGs and RWGs are uniformed distributed along the $y$-axis between $[-\frac{D}{2},\frac{D}{2}]$ m, i.e., the $y$-axis of the $n$-th TWG is $y_{{\sf t},n}=\frac{(2n-1)D}{2N_{\sf t}}-\frac{D}{2}$, the $y$-axis of the $n$-th RWG is $y_{{\sf r},n}=\frac{(2n-1)D}{2N_{\sf r}}-\frac{D}{2}$.
	% Since the PAs on each segment can only be positioned within the segment, we consider that there is only one segment on the TWG, such that each PA can be positioned anywhere on the entire TWG to mitigate the impact of path loss. To improve the capacity of receiving reflected signals, we set four segments on each receiving waveguide.
	There are four segments on each waveguide.  
	The users and targets are uniformly distributed in a $20\times 20$ m$^2$ squared region, i.e., $D=20$ m. The $z$-axis distance between the users/targets and the waveguides is $d=3$ m, $f_c=28$ GHz, $n_{\rm{eff}}=1.4$. The simulation setup is used in the sequel unless otherwise specified.
	
	The GNN is trained with 1,000 samples, which are enough for achieving the best performance according to our evaluations. Since the GNN is trained with unsupervised learning, each training sample only contains the input of the GNN, i.e., the positions of users and the target, which are generated by following the uniform distribution as shown above. The trained GNN is tested on another 100 samples.
	
	Since the Lagrange multiplier method cannot guarantee that the constraints in \eqref{eq:const-r} can be completely satisfied, the learning performance is measured by two metrics, i) \emph{Constraint Satisfication Ratio (CSR)}, which is the ratio of users satisfying the minimum data rate requirement averaged over all the test samples, and ii) \emph{SR}, which is the achieved SR averaged over all the test samples. 
	
	We compare the learning and optimization performance with the following baselines.
	\begin{itemize}
		\item \emph{Random}: The PAs on each waveguide are randomly deployed, and zero-forcing (ZF) is employed as the transmit beamforming.
		\item \emph{CMIMO}: A $N_{\sf t}\times MS$ planar antenna array is used for transmission, and a $N_{\sf r}\times MS$ planar antenna array is used for receiving. The antenna spacing is $\lambda/2$. ZF is still employed for the transmit beamforming.
	\end{itemize}
	In these two baselines, the ZF beamforming can well suppress the MUI but does not take the performance of sensing and the minimal data rate requirement into consideration.
	
	The fine-tuned hyper-parameters of the SWISAC-GNN are shown in Table. \ref{table:hyper-param}. 
	\begin{table}[!htb]
		\centering
		\caption{Hyper-parameters of the SWISAC-GNN}
		\label{table:hyper-param}
		\footnotesize
		\begin{tabular}{c|c|c|c|c}
			\hline\hline
			FNN & \makecell{Number of \\hidden layers} & \makecell{Neurons in\\ each layer} & \makecell{Activation\\ function} & \makecell{Learning\\ rate} \\
			\hline
			Update $x_{{\sf t},msn}^{(\ell+1)}$ & 0* & [] & Sigmoid & \multirow{6}{*}{0.001} \\
			\cline{1-4} 
			Update $x_{{\sf r},ms}^{(\ell+1)}$ & 0 & [] & Sigmoid &  \\
			\cline{1-4} 
			Update $\xi_{nk}^{(\ell+1)}$ & 2 & [8,8] & Tanh &  \\
			\cline{1-4}  
			Update $\beta_{k}^{(\ell+1)}$ & 2 & [8,8] & Softplus & \\
			\cline{1-4} 
			Update $p_k^{(\ell+1)}$ & 2 & [8,8] & Softplus & \\
			\cline{1-4} 
			Update $\mathbf{v}_n^{(\ell+1)}$ & 2 & [8,8] & Softplus & \\
			\hline\hline
		\end{tabular}
		
		*: Zero hidden layer indicates that the FNN is only with an input and an output layer.
	\end{table}
	
	\subsection{Performance Versus the Number of Segments on Each Receiving Waveguide}
	Table \ref{tab:sr-seg} shows how the SR changes with the number of segments on each receiving waveguide, i.e., $S$. When $S=1$, the SWAN reduces to the traditional PASS. The growing trend shown in the figure demonstrates that employing the SWAN architecture can indeed improve the sensing performance. 
	We can also see that the SWISAC-GNN, whose architecture is inspired by the SWISAC-AO algorithm, can achieve better SR than the algorithm. First, a good combiner of the GNN can be learned from data, which helps better convergence after multiple layers (corresponding to the number of iterations of the AO algorithm). Secondly, the number of iterations of the AO algorithm is set as five for affordable computational complexity.
	\begin{table}[!htb]
		\centering
		\footnotesize
		\caption{SR versus number of segments, $N_{\sf t}=16,N_{\sf r}=2, K_{\sf c}=4, K_{\sf s}=2, R_{\rm min}=10$ bps/s.}\label{tab:sr-seg}
		\begin{tabular}{c|c|c|c|c}
			\hline\hline
			Number of segments & 1 & 2 & 3 & 4   \\
			\hline
			SWISAC-GNN & 1.59 & 1.84 & 1.95 & 2.06 \\
			\hline
			SWISAC-AO & 1.47 & 1.76 & 1.82 & 1.97 \\
			\hline\hline
		\end{tabular}
	\end{table}
	%    \begin{figure}[!htb]
	% 	\centering
	% 	\includegraphics[width=.65\linewidth]{Figures/fig-sr-seg.pdf}
	% 	\caption{Sensing performance versus number of segments, $N_{\sf t}=16,N_{\sf r}=2, K_{\sf c}=4, K_{\sf s}=2, R_{\rm min}=10$ bps/s.}
	% 	\label{fig:sr-seg}
	% \end{figure}
	
	\subsection{Performance Under Different Noise Power and Minimum Data Rate Requirement}
\begin{figure}[!htb]
		\centering
		\begin{minipage}[t]{0.85\linewidth}	
			\subfigure[Achieved CSR, $\sigma_{\sf c}^2=-90$ dBm, $\sigma_{\sf s}^2=-120$ dBm.]{
				\includegraphics[width=\textwidth]{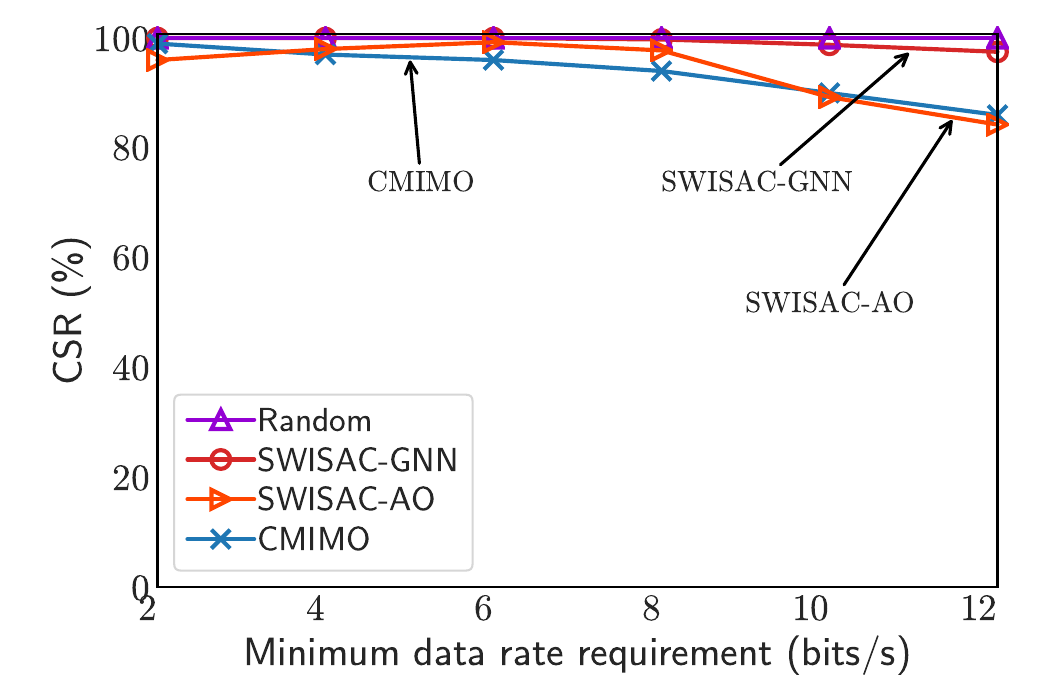}}
		\end{minipage} 
		
		\begin{minipage}[t]{0.85\linewidth}	
			\subfigure[Achieved SR, $\sigma_{\sf c}^2=-90$ dBm, $\sigma_{\sf s}^2=-120$ dBm.]{
				\includegraphics[width=\textwidth]{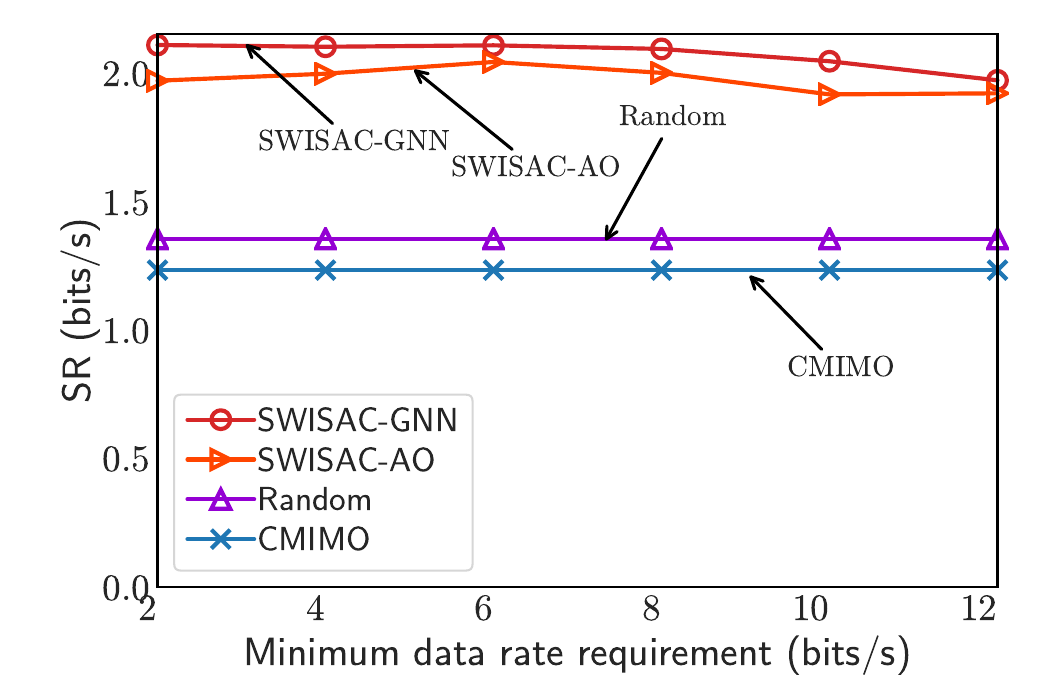}}
		\end{minipage}
		\caption{Performance of beamforming $R_{\rm min}$, $N_{\sf t}=16,N_{\sf r}=2, K_{\sf c}=4, K_{\sf s}=2$.}
		\label{fig:perf-rmin}
	\end{figure}

\begin{figure}[!htb]
		\centering
		\begin{minipage}[t]{0.85\linewidth}	
			\subfigure[Achieved CSR, $R_{\rm min}=10$ bits/s, $\sigma_{\sf s}^2$ is 30 dB lower than $\sigma_{\sf c}^2$.]{
				\includegraphics[width=\textwidth]{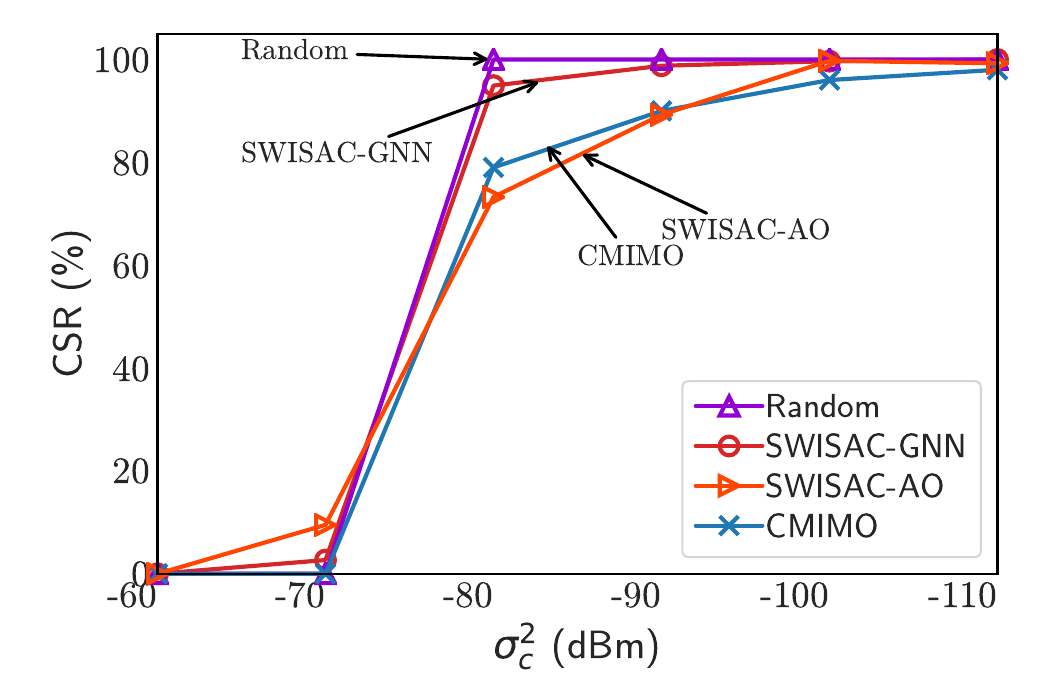}}
		\end{minipage} 
		
		\begin{minipage}[t]{0.85\linewidth}	
			\subfigure[Achieved SR, $R_{\rm min}=10$ bits/s, $\sigma_{\sf s}^2$ is 30 dB lower than $\sigma_{\sf c}^2$.]{
				\includegraphics[width=\textwidth]{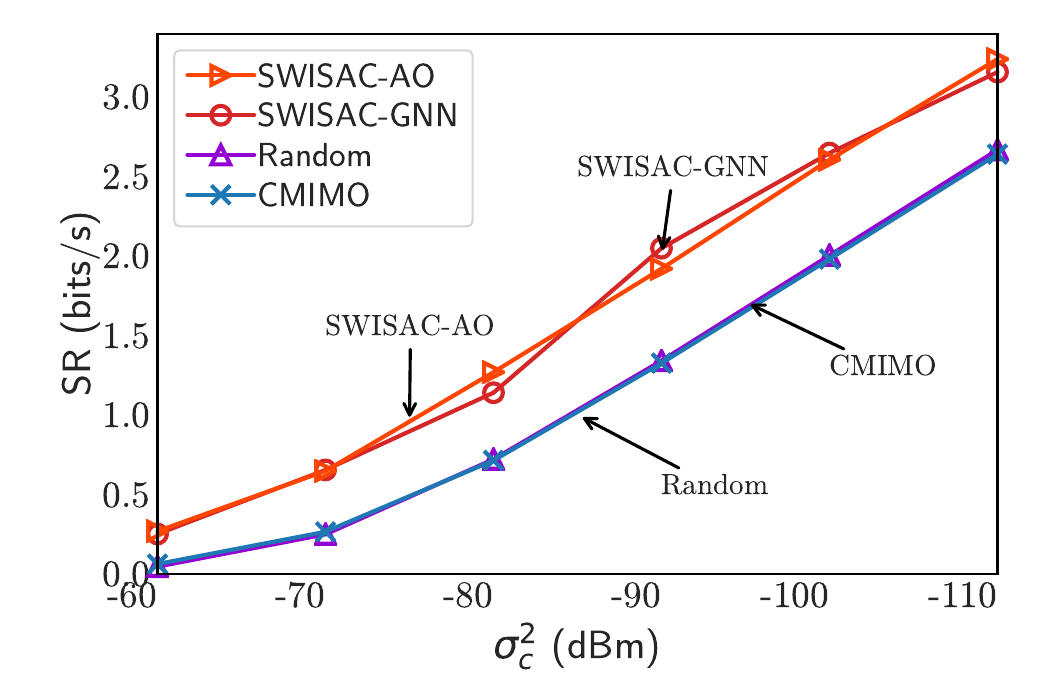}}
		\end{minipage}
		\caption{Performance of beamforming versus noise power, $N_{\sf t}=16,N_{\sf r}=2, K_{\sf c}=4, K_{\sf s}=2$.} \label{fig:perf-snr}
	\end{figure}

	Figs. \ref{fig:perf-rmin} and \ref{fig:perf-snr} show how the CSR and SR change with the noise power and minimum data rate requirement. We can again see that the SWISAC-GNN can achieve SR close to or better than the SWISAC-AO algorithm. Moreover, the CSR of the SWISAC-GNN is much higher than that of the optimization-based method. We can see that the CMIMO and Random methods perform poorly, especially in SR, because they cannot well mitigate the impact of path loss from long-distance transmission, and the sensing performance is not considered in the ZF transmit beamforming.
	
	\subsection{Running Time Comparison}
	In Table \ref{tab:run-time}, we compare the running time of the SWISAC-GNN and the SWISAC-AO algorithm under different problem scales. The running time of CMIMO and Random methods is not provided because they do not perform well. All the running time is obtained on a computer with two Nvidia RTX 4500 GPUs, an Intel Xeon Gold 5148Y CPU with 24 cores and 48 logical processors, and 128 GB RAM.
	
	We can see that the running time of the optimization-based method is in the scale of seconds, and the time increases with the problem scale. This is because the method requires searching the positions of multiple PAs and solving a sub-problem in every iteration, which are time-consuming. By contrast, the running time of GNN is only in the scale of milliseconds. This is because by learning the distribution pattern of PAs and leveraging the optimal solution structure of transmit beamforming, the GNN can directly output the PA positions and transmit beamforming matrix with low complexity of forward propagation. Hence, the GNN can be implemented in real time.
	
	\begin{table*}[!htb]
		\centering
		\caption{Running time averaged over 100 test samples, $\sigma_{\sf c}^2=-90$ dBm, $\sigma_{\sf s}^2=-120$ dBm, $N_{\sf t}=16, N_{\sf r}=2$.}
		\footnotesize
		\label{tab:run-time}
		\begin{tabular}{c|c|cccc|ccc}
			\toprule
			\multirow{2}{*}{\makecell{Problem\\scale}} 
			& \multirow{2}{*}{} 
			& \multicolumn{4}{c|}{$K_s=2$} 
			& \multicolumn{3}{c}{$K_c=4$} \\
			\cmidrule(lr){3-9}
			&  & $K_c=2$ & $K_c=3$ & $K_c=4$ & $K_c=5$ & $K_s=1$ & $K_s=3$ & $K_s=4$ \\
			\midrule
			\multirow{2}{*}{\makecell{Running\\time (s)}} 
			& SWISAC-GNN       & 0.024   & 0.027   & 0.013   & 0.028   & 0.022   & 0.031   & 0.031 \\
			& SWISAC-AO & 2.52  & 5.02   & 6.60     & 9.84  & 5.56  & 6.69   & 6.90 \\
			\bottomrule
		\end{tabular}
	\end{table*}
	
	\subsection{Adaptation to Different Problem Scales}
\begin{figure*}[!htb]
		\centering
		\begin{minipage}[t]{0.3\linewidth}	
			\subfigure[CSR versus $K_{\sf c}$, $K_{\sf s}=2, N_{\sf t}=16, N_{\sf s}=2$]{
				\includegraphics[width=\textwidth]{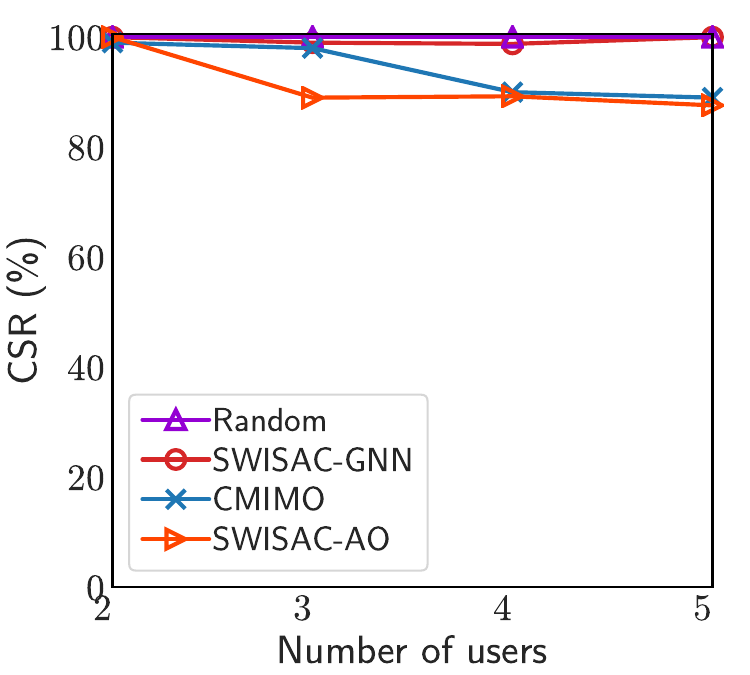}}
		\end{minipage} \hspace{2mm}
		\begin{minipage}[t]{0.3\linewidth}	
			\subfigure[CSR versus $N_{\sf t}$, $K_{\sf s}=2, K_{\sf c}=4, N_{\sf s}=2$]{
				\includegraphics[width=\textwidth]{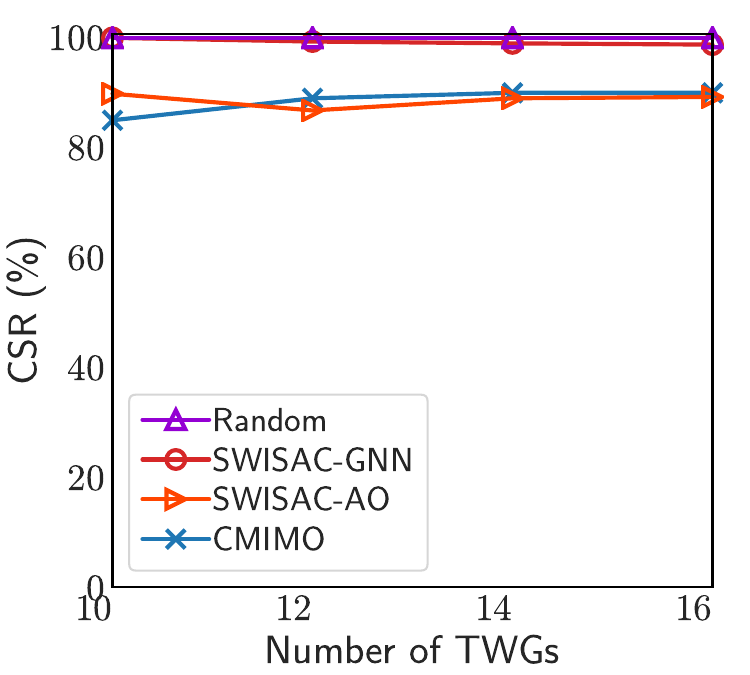}}
		\end{minipage} \hspace{2mm}
		\begin{minipage}[t]{0.3\linewidth}	
			\subfigure[CSR versus $N_{\sf r}$, $K_{\sf s}=2, N_{\sf t}=16, K_{\sf c}=4$]{
				\includegraphics[width=\textwidth]{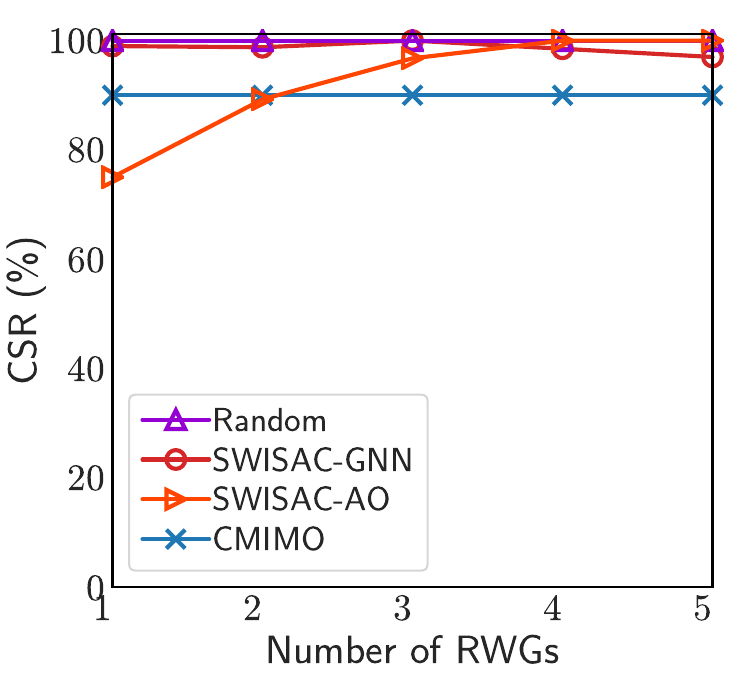}}
		\end{minipage}
		
		\begin{minipage}[t]{0.3\linewidth}	
			\subfigure[SR versus $K_{\sf c}$, $K_{\sf s}=2, N_{\sf t}=16, N_{\sf s}=2$]{
				\includegraphics[width=\textwidth]{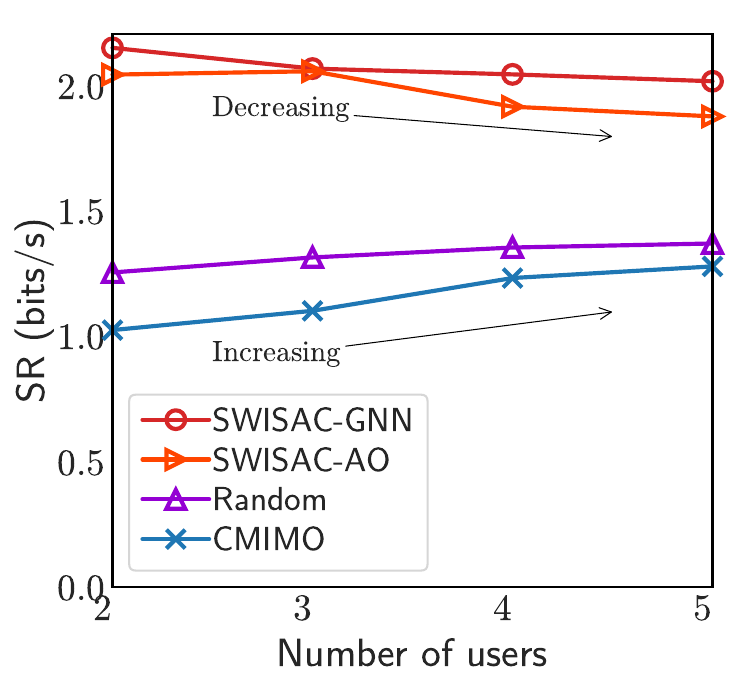}}
		\end{minipage} \hspace{2mm}
		\begin{minipage}[t]{0.3\linewidth}	
			\subfigure[SR versus $N_{\sf t}$, $K_{\sf s}=2, K_{\sf c}=4, N_{\sf s}=2$]{
				\includegraphics[width=\textwidth]{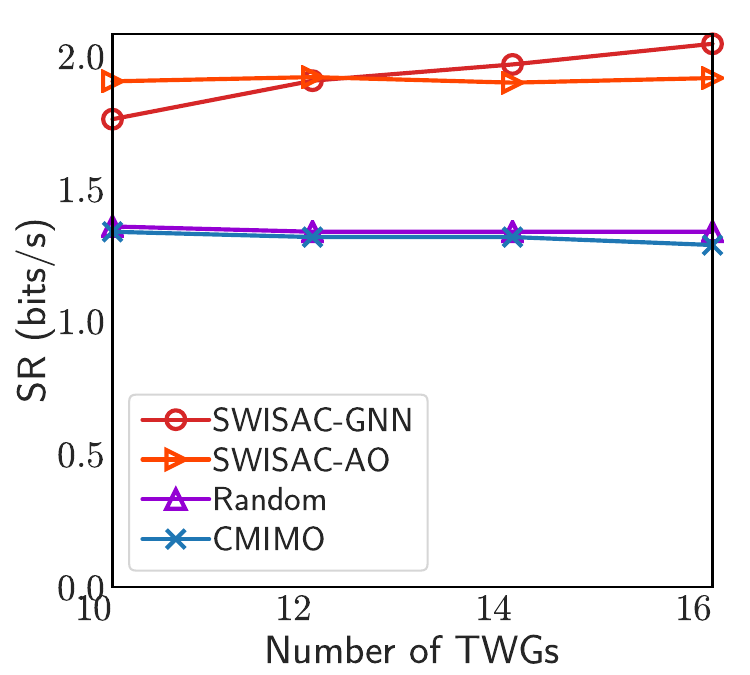}}
		\end{minipage} \hspace{2mm}
		\begin{minipage}[t]{0.3\linewidth}	
			\subfigure[SR versus $N_{\sf r}$, $K_{\sf s}=2, N_{\sf t}=16, K_{\sf c}=4$]{
				\includegraphics[width=\textwidth]{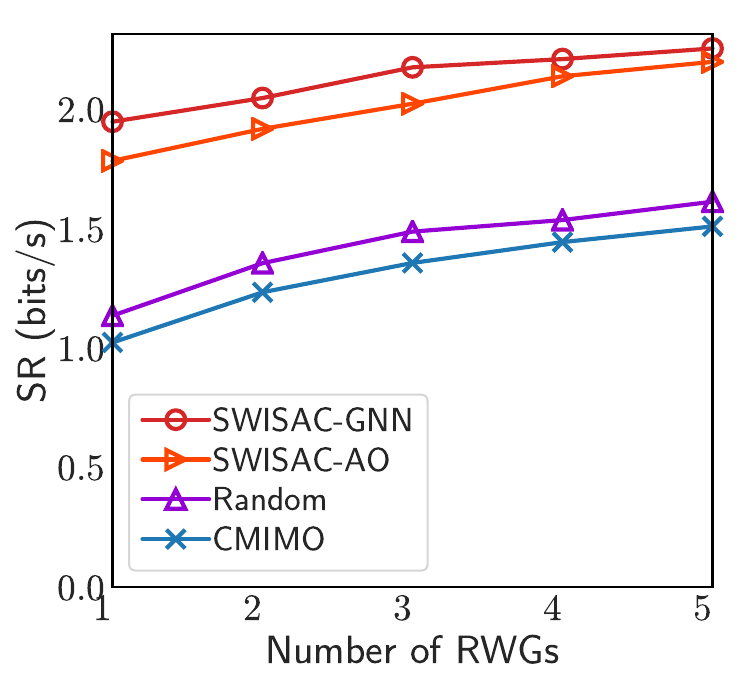}}
		\end{minipage}
		\caption{Performance under different problem scales, $\sigma_{\sf c}^2=-90$ dBm, $\sigma_{\sf s}^2=-120$ dBm, $R_{\min}=10$ bps/s.} \label{fig:adapt}
	\end{figure*}

	We next evaluate the ability of the SWISAC-GNN to fast-adapt to different problem scales. The GNN is trained under the scenario of $K_{\sf c}=4$, $K_{\sf s}=2, N_{\sf t}=16, N_{\sf s}=2$. When the trained GNN is deployed under another problem scale, the GNN is only fine-tuned with three training epochs to adapt to the new scenario. As can be seen from Fig. \ref{fig:adapt},  the GNN can still achieve high SR in all the problem scales, while the CSR of the GNN is much higher than the SWISAC-AO algorithm and other baselines. This demonstrates the fast adaptation ability of the SWISAC-GNN in scenarios where the problem scale varies frequently.
	
	In subfigure (d), we can see that the SR of SWISAC-GNN and SWISAC-AO slightly degrades with the number of users, while the SR of the two baselines increases with $K_{\sf c}$, which seems counterintuitive. This is because the transmit beamforming optimized by the proposed algorithm and GNN takes the QoS of users into consideration. With the increase in the number of users, the direction and power of beamforming vectors should be biased to the users to guarantee the QoS requirements, hence the sensing performance degrades. For the two baselines, ZF beamforming is adopted, such that the beamforming lies in the subspace spanned by the channel vectors of users. With the increase of the users, the probability that the channel vector of the sensing target also lies in the user channel subspace grows, such that the sensing performance is better.

	\section{Conclusions} \label{sec:conclusion}
	We proposed a SWISAC-AO algorithm and a SWISAC-GNN architecture to optimize and learn beamforming in a PASS-aided UAV system for ISAC, aiming to maximize the rate of sensing multiple targets while satisfying minimal data rate requirements of multiple users. The algorithm was first proposed based on AO, where the optimal solution structure of transmit beamforming was derived. Then, the SWISAC-GNN architecture was proposed, where the alternative update procedure and information aggregation mechanism were inspired by the  SWISAC-AO algorithm. The GNN satisfies the permutation property of the beamforming policy. Simulation results show that the SWISAC-GNN can well-satisfy the minimal data rate constraints while achieving SR close to or higher than the SWISAC-AO algorithm. The GNN is also with much lower inference time than the algorithm, and can be fast-adapted to different problem scales.

	%In this paper, we proposed a GNN architecture to jointly learn pinching and transmit beamforming in PASS-aided ISAC system, aiming to maximize sensing rate while satisfying minimal data rate requirements. The GNN firstly learn the pinching beamforming over a heterogeneous graph, by incorporating mathematical models to update representations of edges and vertices and then iteratively updating their actions. After the pinching beamforming is known, the optimal solution structure of transmit beamforming in a special channel orthogonal condition is derived. We incorporate this structure with the GNN to learn transmit beamforming. Simulation results show that the SWISAC-GNN can well-satisfy the minimal data rate constraints while achieving higher SR than the baseline methods with low inference time. The generalizability of the GNN, as well as a more general case with multiple sensing targets will be investigated in future works. 
	
	\begin{appendices} \numberwithin{equation}{section}
		\renewcommand{\thesectiondis}[2]{\Alph{section}}
		\section{Proof of Proposition \ref{prop:structure}}\label{appendix:structure}
		The Lagrangian function of problem \eqref{eq:opt-tb3} can be expressed as,
		\begin{align}\label{eq:lagrange}
			&\tilde{L}(\mathbf{W},\mathbf{V},\bm{\Theta},\bm{\beta}) = L(\bm{\Phi}_{\sf t}, \bm{\Phi}_{\sf r},\mathbf{W},\mathbf{V},\mathbf{{\bm\Theta}}) - \notag\\&\sum_{k=1}^{K_{\sf c}}\beta_k \Big(\frac{1}{\gamma\sigma_{\sf c}^2} |\tilde{\mathbf{h}}_k^{\sf H}\mathbf{w}_k|^2-\frac{1}{\sigma_{\sf c}^2}\sum_{j=1,j\neq k}^{K_{\sf c}} |\tilde{\mathbf{h}}_k^{\sf H}\mathbf{w}_j|^2 - 1\Big) + \notag\\
			&\nu (\|\mathbf{W}\|_F^2-P_{\max}),
		\end{align}
		where $\nu\geq0, \beta_k\geq 0,k=1,\cdots,K_{\sf c}$ are the Lagrange multipliers. 
		
		Finding the stationary Karush–Kuhn–Tucker (KKT) point from $\partial\tilde{L}/\partial \mathbf{w}_k=0$ yields,
		\begin{align}
			&\Bigg(\sum_{n=1}^{N_{\sf r}} (\mathbf{v}_n^{\sf H}\bm{\Theta}_n\mathbf{v}_n)\tilde{\mathbf{h}}_{{\sf s},n}\tilde{\mathbf{h}}_{{\sf s},n}^{\sf H}\Bigg)\mathbf{w}_k - \sum_{n=1}^{N_{\sf r}} \tilde{\mathbf{h}}_{{\sf s},n} (\bm{\Theta}_n\mathbf{v}_n)_k - \notag\\&\frac{\beta_k}{\gamma\sigma_{\sf c}^2}(\tilde{\bf h}_k\tilde{\bf h}_k^{\sf H})\mathbf{w}_k + \sum_{j=1}^{K_{\sf c}} \frac{\beta_j}{\sigma_{\sf c}^2}\tilde{\mathbf{h}}_j\tilde{\bf h}_j^{\sf H}\mathbf{w}_k + \mu\mathbf{w}_k = 0,
		\end{align}
		where $(\mathbf{x})_k$ denotes the $k$-th element in $\mathbf{x}$. Then, we have,
		\begin{align}
			&\Bigg(\sum_{n=1}^{N_{\sf r}} (\mathbf{v}_n^{\sf H}\bm{\Theta}_n\mathbf{v}_n)\tilde{\mathbf{h}}_{{\sf s},n}\tilde{\mathbf{h}}_{{\sf s},n}^{\sf H}+\sum_{j=1}^{K_{\sf c}} \frac{\beta_j}{\sigma_{\sf c}^2}\tilde{\mathbf{h}}_j\tilde{\bf h}_j^{\sf H} +\nu\mathbf{I}\Bigg)\mathbf{w}_k =\notag\\& \frac{(1+\gamma)\beta_k}{\gamma\sigma_{\sf c}^2}(\tilde{\bf h}_k\tilde{\bf h}_k^{\sf H})\mathbf{w}_k + \sum_{n=1}^{N_{\sf r}} \tilde{\mathbf{h}}_{{\sf s},n} (\bm{\Theta}_n\mathbf{v}_n)_k \\
			& \mathbf{w}_k = \Bigg(\sum_{n=1}^{N_{\sf r}} (\mathbf{v}_n^{\sf H}\bm{\Theta}_n\mathbf{v}_n)\tilde{\mathbf{h}}_{{\sf s},n}\tilde{\mathbf{h}}_{{\sf s},n}^{\sf H}+\sum_{j=1}^{K_{\sf c}} \frac{\beta_j}{\sigma_{\sf c}^2}\tilde{\mathbf{h}}_j\tilde{\bf h}_j^{\sf H}+\nu\mathbf{I}\Bigg)^{-1} \notag\\& \Bigg(\frac{(1+\gamma)\beta_k}{\gamma\sigma_{\sf c}^2}(\tilde{\bf h}_k^{\sf H}\mathbf{w}_k)\tilde{\bf h}_k + \sum_{n=1}^{N_{\sf r}}  (\bm{\Theta}_n\mathbf{v}_n)_k \tilde{\mathbf{h}}_{{\sf s},n}\Bigg).
		\end{align}
		
		Furthermore, it is not hard to prove that the problem \eqref{eq:opt-tb3} always achieves optimality when $\|\mathbf{W}\|_F^2=P_{\max}$. Then, the optimal transmit beamforming vector can be expressed as \eqref{eq:structure}.
		\section{Proof of Proposition \ref{prop:pe}}\label{appendix}
		We start by proving that the input-output relation of each layer of the GNN can satisfy the permutation property. Then, by stacking $L$ layers with the same update equation, the input-output relation of the GNN can satisfy the permutation property.
		
		We can see from Fig. \ref{fig:gnn} that the input-output relation in each layer is composed by several functions, i.e., $F_{\sf t}(\cdot), F_{\sf r}(\cdot), F_{\bm\beta}(\cdot), F_{\bm\Xi}(\cdot), F_{\mathbf{p}}(\cdot), F_{\mathbf{V}}(\cdot)$. We next prove that every function satisfies the permutation property. Specifically, we take $F_{\bm\beta}(\cdot)$ as an example for illustration, and the proofs for other functions are similar.
		
		By changing the order of segments of transmit waveguides with $\bm\Omega_{\sf t}$, and changing the orders of transmit waveguides and users with $\bm\Pi_{\sf t}$ and $\bm\Pi_{\sf ue}$, the input of $F_{\bm\beta}(\cdot)$ becomes $\bm\Pi_{\sf ue}^{\sf T}\bm\beta^{(\ell)}=[\beta_{\pi_{\sf ue}(1)},\cdots,\beta_{\pi_{\sf ue}(K)}]^{\sf T}\footnote{$\pi_{\sf ue}(k)$ denotes the index permuted from index $k$ with $\bm\Pi_{\sf ue}$, i.e., $\bm\Pi_{\sf ue}^{\sf T}[1,2,3]^{\sf T}=[\pi_{\sf ue}(1),\pi_{\sf ue}(2),\pi_{\sf ue}(3)]^{\sf T}$.},\bm\Pi_{\sf t}^{\sf T}\mathbf{W}^{(\ell)}\bm\Pi_{\sf ue},\bm\Phi_{\sf t}^{(\ell)}\bm\Omega_{\sf t},\mathbf{U}\bm\Pi_{\sf ue}$. In this case, the $k$-th element in the output vector of $F_{\bm\beta}(\cdot)$ becomes,
		\begin{align}\label{eq:perm}
			&\mathsf{FNN}\Big(\beta_{\pi_{\sf ue}(k)}^{(\ell)}, \mathbf{h}^{\sf H}(\mathbf{u}_{\pi_{\sf ue}(k)}, \bm\Phi_{\sf t}^{(\ell)}\bm\Omega_{\sf t})\mathbf{G}(\bm\Phi_{\sf t}^{(\ell)}\bm\Omega_{\sf t})\bm\Pi_{\sf t}^{\sf T}\cdot\notag\\
			&\hspace{10mm}\mathbf{W}^{(\ell)}\bm\Pi_{\sf ue}\bm\Pi_{\sf ue}^{\sf T}\mathbf{b}_{\pi_{\sf ue}(k)}^{(\ell)}\Big)\notag\\
			&\overset{(a)}{=}\mathsf{FNN}\Big(\beta_{\pi_{\sf ue}(k)}^{(\ell)}, \mathbf{h}^{\sf H}(\mathbf{u}_{\pi_{\sf ue}(k)}, \bm\Phi_{\sf t}^{(\ell)})\bm\Omega_{\sf t}\bm\Omega_{\sf t}^{\sf T}\mathbf{G}(\bm\Phi_{\sf t}^{(\ell)})\bm\Pi_{\sf t}\bm\Pi_{\sf t}^{\sf T}\notag\\
			&\hspace{10mm}\mathbf{W}^{(\ell)}\bm\Pi_{\sf ue}\bm\Pi_{\sf ue}^{\sf T}\mathbf{b}_{\pi_{\sf ue}(k)}^{(\ell)}\Big)\notag\\
			&\overset{(b)}{=}\mathsf{FNN}\Big(\beta_{\pi_{\sf ue}(k)}^{(\ell)}, \mathbf{h}^{\sf H}(\mathbf{u}_{\pi_{\sf ue}(k)}, \bm\Phi_{\sf t}^{(\ell)})\mathbf{G}(\bm\Phi_{\sf t}^{(\ell)})\mathbf{W}^{(\ell)}\mathbf{b}_{\pi_{\sf ue}(k)}^{(\ell)}\Big)\notag \\
			&=\beta_{\pi_{\sf ue}(k)}^{(\ell+1)},
		\end{align}
		where $(a)$ comes from $\mathbf{h}(\mathbf{u}_{\pi_{\sf ue}(k)}, \bm\Phi_{\sf t}^{(\ell)}\bm\Omega_{\sf t})=\bm\Omega_{\sf t}^{\sf T}\mathbf{h}(\mathbf{u}_{\pi_{\sf ue}(k)}, $ $\bm\Phi_{\sf t}^{(\ell)})$ and $\mathbf{G}(\bm\Phi_{\sf t}^{(\ell)}\bm\Omega_{\sf t})=\bm\Omega_{\sf t}^{\sf T}\mathbf{G}(\bm\Phi_{\sf t}^{(\ell)})\bm\Pi_{\sf t}$, which can be obtained by substituting the permuted variables into the expressions of free-space channel vector and in-waveguide channel matrix, $(b)$ comes from $\bm\Pi\bm\Pi^{\sf T}=\mathbf{I}$ for any permutation matrix $\bm\Pi$. From \eqref{eq:perm}, we can obtain that the output of $F_{\bm\beta}(\cdot)$ becomes $\bm\Pi_{\sf ue}^{\sf T}\bm\beta$. Then, $\bm\Pi_{\sf ue}^{\sf T}\bm\beta=F_{\bm\beta}(\bm\Pi_{\sf ue}^{\sf T}\bm\beta^{(\ell)},\bm\Pi_{\sf t}^{\sf T}\mathbf{W}^{(\ell)}\bm\Pi_{\sf ue},\bm\Phi_{\sf t}^{(\ell)}\bm\Omega_{\sf t},\mathbf{U}\bm\Pi_{\sf ue})$ holds, which indicates that only the order of elements in the output vectors is permuted while the values keep unchanged. 
		
		Similarly, we can prove that $F_{\sf t}(\cdot), F_{\sf r}(\cdot), F_{\bm\Xi}(\cdot), F_{\mathbf{p}}(\cdot), $ $F_{\mathbf{V}}(\cdot)$ also satisfies the permutation property. Denote the input-output relation of the $\ell$-th layer of the GNN as $\{\bm\Phi_{\sf t}^{(\ell+1)},\bm\Phi_{\sf r}^{(\ell+1)},\mathbf{W}^{(\ell+1)},\mathbf{V}^{(\ell+1)}\}=G(\bm\Phi_{\sf t}^{(\ell)},\bm\Phi_{\sf r}^{(\ell)},\mathbf{W}^{(\ell)},\mathbf{V}^{(\ell)},\mathbf{U}_{\sf tar},\mathbf{U})$. Since all the functions in the $\ell$-th layer satisfy the permutation property, it is not hard to prove that $G(\cdot)$ satisfies
		\begin{align}
			&\{\bm\Phi_{\sf t}^{(\ell+1)}\bm\Omega_{\sf t},\bm\Phi_{\sf r}^{(\ell+1)}\bm\Omega_{\sf r},\bm\Pi_{\sf t}^{\sf T}\mathbf{W}^{(\ell+1)}\bm\Pi_{\sf ue},\bm\Pi_{\sf r}^{\sf T}\mathbf{V}^{(\ell+1)}\bm\Pi_{\sf ue}\}=\notag\\
			&G(\bm\Phi_{\sf t}^{(\ell)}\bm\Omega_{\sf t},\bm\Phi_{\sf r}^{(\ell)}\bm\Omega_{\sf r},\bm\Pi_{\sf t}^{\sf T}\mathbf{W}^{(\ell)}\bm\Pi_{\sf ue},\bm\Pi_{\sf r}^{\sf T}\mathbf{V}^{(\ell)}\bm\Pi_{\sf ue},\mathbf{U}_{\sf tar}\bm\Pi_{\sf tar},\mathbf{U}\bm\Pi_{\sf ue}),\notag
		\end{align}
		i.e., the input-output relation is not affected by changing the orders of segments, waveguides, users and targets. Then, by stacking $L$ layers with the same structure, it is not hard to prove that the input-output relation of the GNN satisfies the permutation property in \eqref{eq:pe}.
	\end{appendices}

	\bibliography{IEEEabrv,GJ}
\end{document}